\documentclass{optica-article}

\journal{opticajournal} 

\articletype{Research Article}

\usepackage{bm}
\usepackage{booktabs}

\renewenvironment{abstract}
  {\vskip1pc\noindent\textbf{Abstract:\space}}
  {\\[12pt]\par\vskip12pt}

\begin{document}

\title{Non-contact lens metrology from optical flow and differentiable ray tracing}

\author{Hector A. Jimenez-Romero,\authormark{1} and Wolfgang Heidrich \authormark{1,*}}

\address{\authormark{1}King Abdullah University of Science and Technology (KAUST), Thuwal 23955-6900, Saudi Arabia
}

\email{\authormark{*}wolfgang.heidrich@kaust.edu.sa} 
\begin{abstract}
Accurate lens surface metrology is essential to determine the manufactured geometry of refractive optical elements and their deviation from the specified design. We demonstrate a transmission metrology method that reconstructs refractive lens surfaces from the displacement of a wavelet-noise-patterned mask. Lens-induced distortion is represented by a dense displacement field. A differentiable ray tracer fits the lens geometry using a calibrated near-telecentric camera model with an effective finite ray origin. We validate the method on commercial lenses. The recovered geometries agree with nominal specifications, and the plano-aspheric reconstruction is compared with chromatic confocal profilometry.\\
\end{abstract}

\section{Introduction}

Surface-form errors have an impact on the optical power, the amount of aberration, and the rate at which spherical, aspheric, and freeform components can be manufactured \cite{elhayek2014,ye2021,hao2023,mishra2024,Rolland:21}. Although tactile coordinate measuring machines are capable of taking accurate surface measurements, the process is relatively slow and physical contact might cause damage to the optical element \cite{Petz2009-yd,elhayek2014,beutler2016,ramirez2021,mishra2024,miao2026}. Non-contact interferometric techniques offer high spatial resolution, but their dynamic range can be limited when dealing with surfaces that have large departures, in which case special null optics or computer-generated holograms may be necessary \cite{beutler2016,miao2026,mishra2024,wyant1974}. Deflectometric methods determine the surface shape by measuring the deviations of rays and are able to handle larger surface slopes \cite{diaz2000,Petz2009-yd,knauer2004,huerta2021}. With respect to refractive optics, transmission measurements continue to be difficult since the displacement of the observed rays depends on the geometry of the refracting surfaces, the refractive index, and the imaging setup \cite{huerta2021,ye2021,wang2021,burke2023}.

Pattern-based transmission measurements offer a direct means of observing these ray deviations. Both background-oriented Schlieren (BOS) and synthetic Schlieren methods estimate refraction by means of the apparent displacement of a known background and have been widely applied in the measurement of transparent media and liquid interfaces \cite{Raffel2015,Moisy2009}. The use of optical-flow processing with random-dot and wavelet-noise backgrounds allows for dense estimates of displacement over a wide range of displacements \cite{Cakir2022}, and telecentric imaging has been employed to enhance the spatial fidelity of such measurements \cite{Ota2015}. Early pattern-transmission techniques for optical surfaces also determined the geometry by fitting surface models to the measured ray deflections \cite{Petz2009-yd,Fischer2014}. The methods mentioned above demonstrate the sensing principle employed here: refraction is observed via the displacement of a known pattern, and the displacement field that is measured serves as the data for surface reconstruction.

Recent computational techniques have made it more practical to carry out inverse reconstruction based on measurements of ray deflection. Differentiable refractive deflectometry simultaneously determines the shape and position of the lens by means of a differentiable forward model \cite{wang2021}. The refractive null-screen methods have managed to recover plano-convex aspheres by iteratively optimizing the distortions of the measured pattern \cite{huerta2024}. Vision Ray Metrology is able to reconstruct surface topography directly from the measured ray deflections and has been used for form and mid-spatial-frequency characterization \cite{ramirez2021,ramirez2024}. These methods also highlight the importance of calibration, telecentricity, axial positioning, and multi-surface conditioning in transmission metrology \cite{ramirez2021,ramirez2024,Xu:2019}.

We reconstruct refractive lens surfaces from the displacement of a fixed wavelet-noise mask as seen through the test optic. An image of the mask taken as a reference and a second image obtained with the lens together give a dense displacement field. This field is first set up using a physical radial model which has been fitted to a few image correspondences and is then improved by means of direct image-consistency optimization. A differentiable ray tracer is then used to estimate the lens shape, employing a calibrated near-telecentric camera model that has an effective finite ray origin. For spherical lenses and exploratory residual analyses, reconstruction proceeds from a conic surface to a dense heightmap and final parametric fitting. The quantitative AL2550 benchmark directly optimizes a shared rotational asphere with a separate pose for each acquisition. The method is tested on commercial plano-convex, biconvex, and plano-aspheric lenses, and the reconstructed aspheric surface is compared with that obtained by chromatic confocal profilometry. The residual heightmaps are analyzed in a Zernike basis in order to characterize the low-order surface structure beyond the fitted nominal geometry.

\section{Methods}

The process begins with two images of the chrome mask and a lens-induced displacement field fitted by differentiable ray tracing. The spherical-lens workflow then reduces an optimized heightmap to optical shape parameters and residual maps; the AL2550 benchmark directly fits rotational-asphere parameters. Unless otherwise stated, the Methods figures refer to the Thorlabs LA1608-A plano-convex lens as the example under consideration. This measurement demonstrates distortion retrieval, differentiable surface reconstruction, and residual-based aberration analysis.

\subsection{Experimental setup}

\begin{figure}[htbp]
\centering
\includegraphics[width=\linewidth]{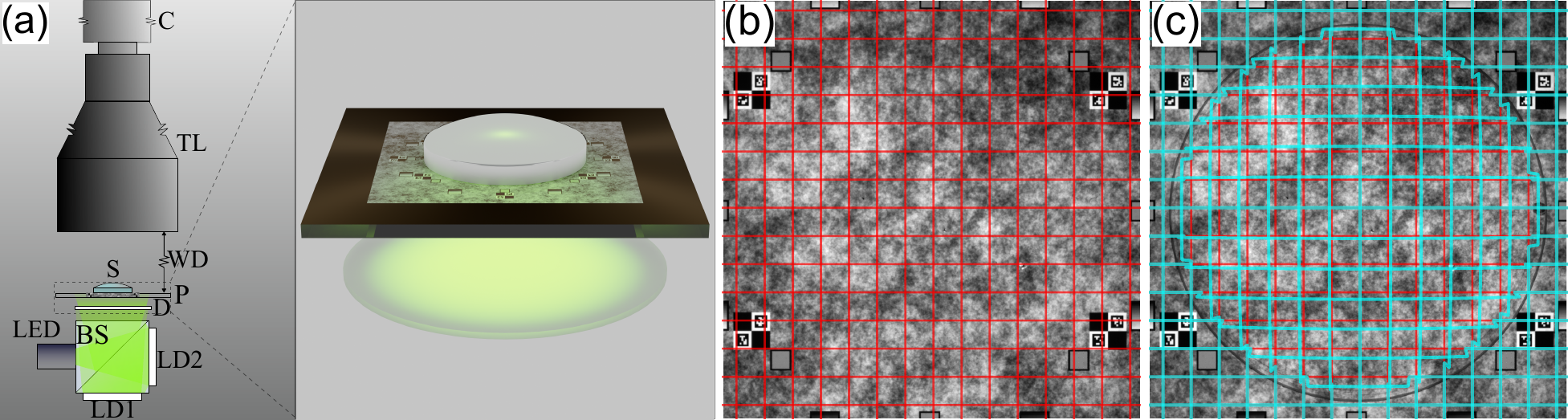}
\caption{Experimental setup and data capture for lens metrology. (a) The imaging system comprises a FLIR camera (C, GS3-U3-41C6M-C) and a telecentric lens (TL, Edmund Optics 3/4 in C-Mount TitanTL\textregistered{} series) having a working distance of $\mathrm{WD} = 110 \, \text{mm}$ and a nominal magnification of $M = 0.184\times$. The lens sample (S) is positioned on the chrome mask, which has the pattern (P). The illumination is provided by a Thorlabs M565L3 LED, with a dominant wavelength of $\lambda = 565 \, \text{nm}$, and is directed through a 50:50 beam splitter (BS) and two matte near-Lambertian diffusers (LD1, LD2). The inset shows the back-illumination coming out of the diffuser, the chrome binary mask, and the lens placed on top of the mask. On the right side, there are examples of the data obtained, illustrating (b) the reference pattern and (c) the pattern distorted by the lens, together with overlaid grids that track the distortion field.}
\label{fig:experimental_setup}
\end{figure}

Each measurement uses a reference image of the undistorted pattern and a second image with the lens resting on the mask. Figure~\ref{fig:experimental_setup} shows the setup. The vertical configuration lets gravity hold the sample and mask in contact, avoiding the mechanical alignment fixtures required in horizontal arrangements and keeping the full aperture visible. Because the edged pattern is exposed directly to the lens back surface, no cover glass is introduced between the sample and target \cite{MAESTRO-WATSON2017, burke2023,ramirez2024}.

The optical system consists of three components: background illumination at the bottom, a binary-pattern measurement plate in the middle, and a telecentric imaging system above it. In an ideal object-space telecentric system, chief rays are parallel to the optical axis, reducing perspective distortion and supporting an orthographic camera approximation. Each image point still receives a finite cone of rays set by the numerical aperture. In practice, telecentric lenses differ slightly from this ideal; the calibration method employed to estimate this deviation and incorporate it into the ray model is given in the Supplementary Material (Section~\textit{Imaging-System Calibration}).

\subsubsection{Measuring pattern}
A number of targets are capable of detecting transmission distortion in a refractive optic, such as ground-glass diffusers, printed transparencies, laser-printed random-dot targets employed in synthetic Schlieren, and etched binary masks \cite{Moisy2009,Berto2017,Wang2017,Wang2019}. We have chosen a chrome-etched binary wavelet-noise mask for two reasons. First, the binary chrome pattern provides high contrast and can be manufactured on a scale smaller than that of the camera's sampling; the feature size of the $1.975\,\mu\text{m}$ mask, when imaged through the 0.184X telecentric lens, corresponds to about 15 mask pixels being mapped onto each camera pixel. Secondly, wavelet noise is well suited to optical flow since its scale-invariant histogram distribution is compatible with the coarse-to-fine pyramids employed by classic flow algorithms, and previous studies related to BOS have directly evaluated wavelet-noise backgrounds using optical-flow processing \cite{Cook2005,RN592,Farneback2003,WANG2024102029,Cakir2022}.

Details regarding pattern fabrication and image-domain validation can be found in the Supplementary Material (Section~\textit{Pattern Design and Capture Fidelity}).

\subsection{Distortion field retrieval}\label{sec:distortion_retrieval}
The measurement of the distortion field is connected with wavefront-sensing techniques that use random patterns \cite{Berto2017, Wang2017, Wang2019}. These methods compare a reference image with an image of a shifted pattern and, in the paraxial regime, recover phase by optimizing a potential field. When carrying out measurements with a clear-aperture lens, large ray deviations near the aperture edge can challenge the paraxial approximation. We begin by using a simple geometrical model to initialize the displacement field and then improve it through direct image-consistency optimization in such a way that the warped image stays in agreement with the measured data. The physical radial model employed for this initial step is explained in the Supplementary Material (Section~\textit{Initial State Estimation of the Displacement Field}). The definition of the loss function, the regularization terms, and the details of the optimization used in the dense refinement are provided in the same section, in \textit{Differentiable optimization refinement}.

Instead of using physical radial initialization, the initial field can be calculated using a general optical-flow method. We used OpenCV's Farnebäck implementation \cite{FarnebackOpenCV, Farneback2003}. It produces a dense displacement map, but its accuracy decreases as surface slope and displacement increase.

\begin{figure}[htbp]
\centering
\includegraphics[width=0.75\linewidth]{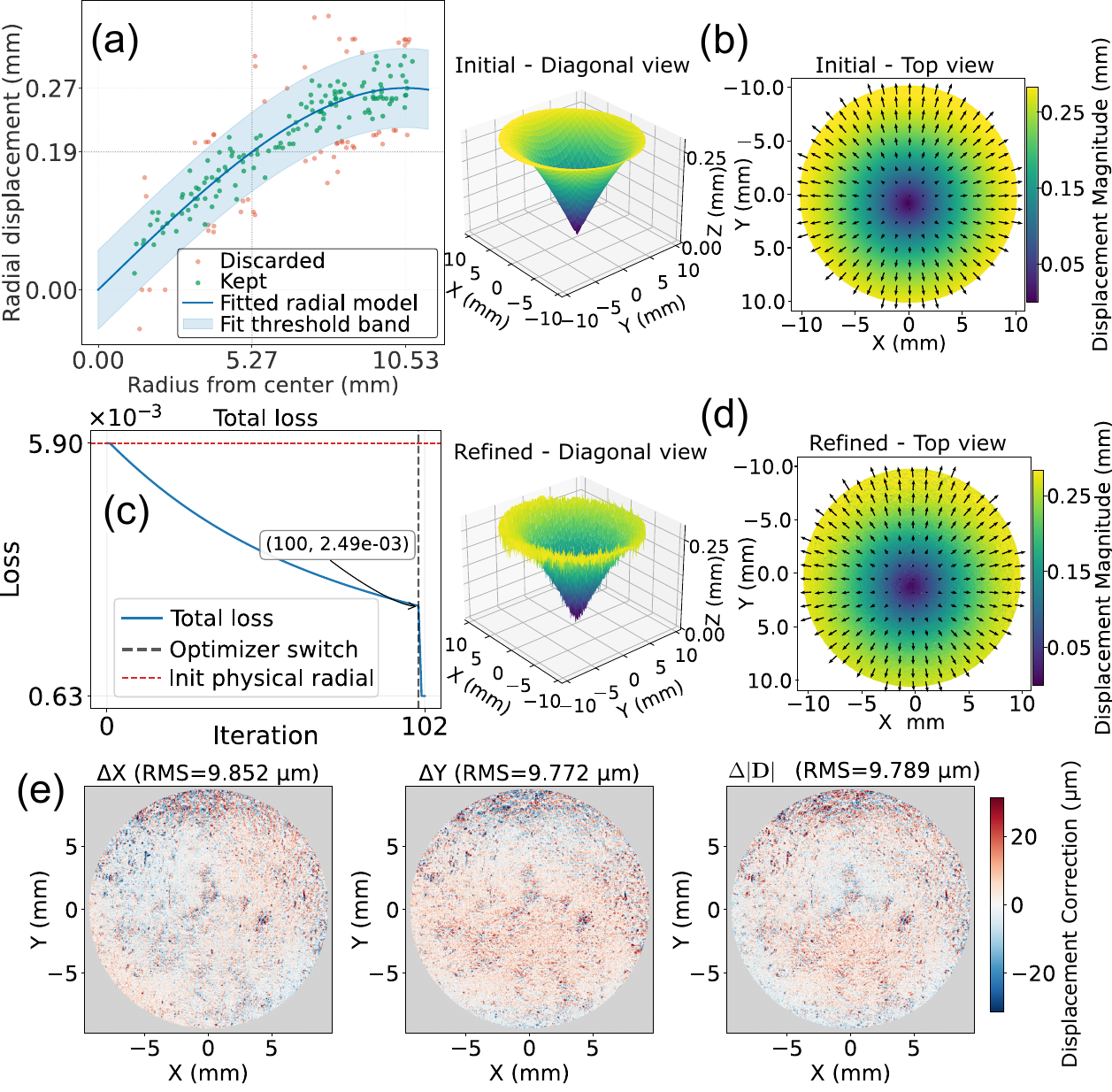}
\caption{Dense displacement-field retrieval and refinement.
(a) The physical radial initialization is obtained by fitting the sparse feature correspondences; the matches that are accepted and those that are discarded are displayed together with the fitted radial model and its threshold band.
(b) The initial displacement field is shown as a three-dimensional magnitude map and a top-view vector field.
(c) The loss evolves during the process of dense image-consistency refinement, with the optimizer change occurring at iteration 100.
(d) The refined displacement field is shown using the same representations as in (b).
(e) Displacement corrections introduced by refinement, \(\mathbf{D}_{\mathrm{ref}}-\mathbf{D}_{\mathrm{init}}\), shown for \(\Delta X\), \(\Delta Y\), and \(\Delta|\mathbf{D}|\). RMS values are reported for each map, with the shared correction scale given in micrometers.}
\label{fig:flow_field}
\end{figure}

A representative measurement is illustrated in Figure~\ref{fig:flow_field} in terms of the physical initialization and the following dense refinement; an example of the pattern validation can be found in the Supplementary Material (Section~\textit{Pattern Design and Capture Fidelity}).

\subsection{Differentiable surface solver}
\begin{figure}[htbp]
\centering
\includegraphics[width=\linewidth]{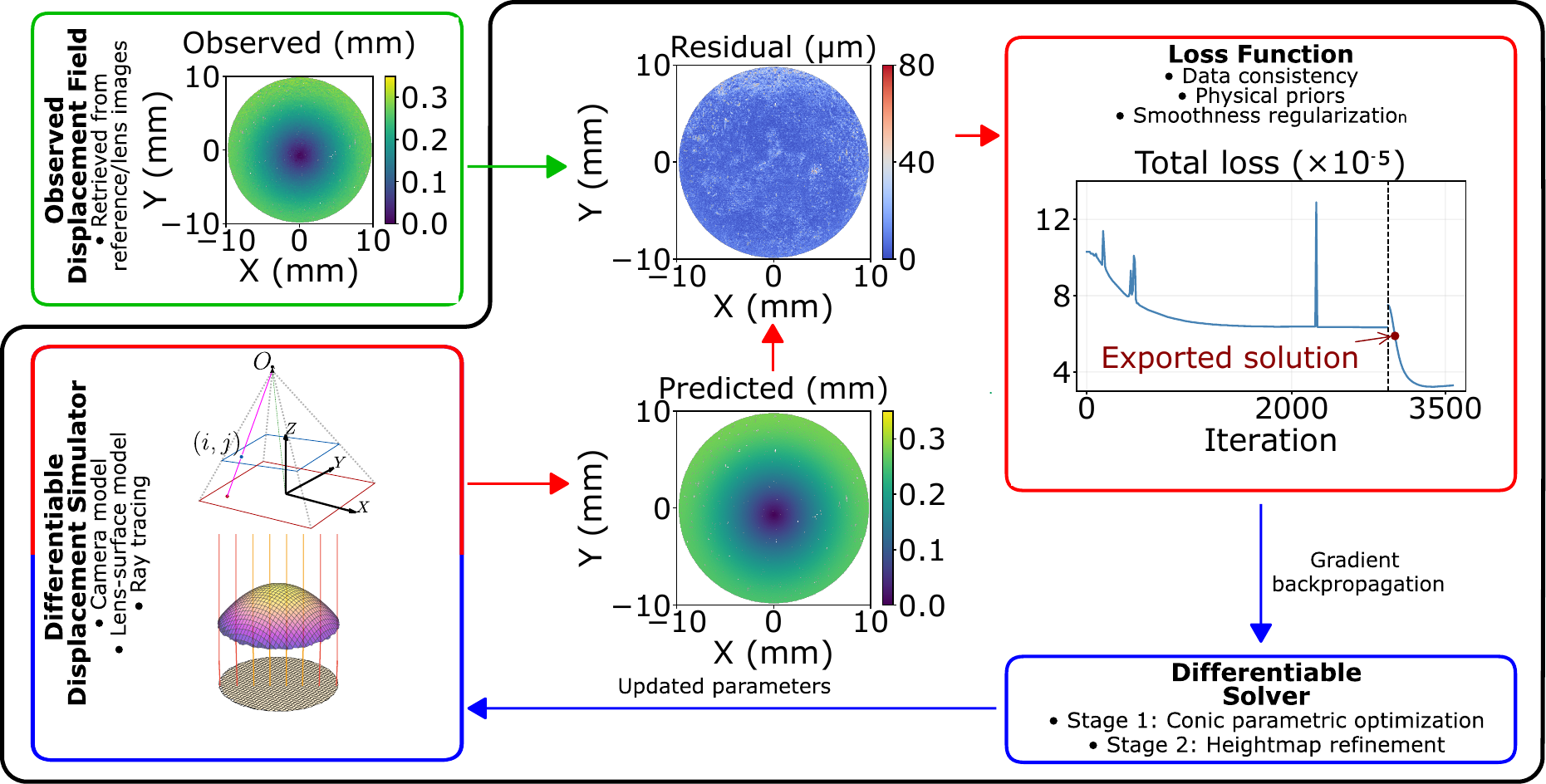}
\caption{Differentiable optimization pipeline for the spherical-lens heightmap workflow, illustrated with LA1608-A. Observed and predicted sensor-displacement magnitudes share a color scale; the residual is the magnitude of their vector difference. All maps show the same valid measurement region in a $20\times20\,\mathrm{mm}$ field. The objective combines data consistency, physical constraints, and heightmap smoothness. Gradients update the reconstruction parameters through the forward model. The sampled loss history marks the conic-to-heightmap transition (dashed line). The dot marks the exported heightmap at its original search iteration: the lowest-objective evaluated state passing physical-ray and stable-inversion checks. The AL2550 rotational benchmark uses the separate direct parametric protocol described in the text.}
\label{fig:displacement_comparison}
\end{figure}

Figure~\ref{fig:displacement_comparison} gives an overview of the reconstruction loop. In each iteration, the present lens and camera parameters produce a predicted displacement field using differentiable ray tracing. This predicted field is then compared with the observed field obtained in Section~\ref{sec:distortion_retrieval}, and the displacement residual from this comparison is used in the data-consistency term. The reconstruction is subject to physical and geometric penalties, and a weak smoothness regularization is applied when refining the heightmap. The gradients are backpropagated through the differentiable forward model in order to update the reconstruction parameters. The complete definition of the loss function is provided in the Supplementary Material (Section~\textit{Reconstruction Objective and Optimization Stages}).

During reconstruction, we optimize lens parameters together with auxiliary alignment parameters that primarily affect observability, including lens decenter in the pattern plane $(x,y)$ and a small angular pose correction for the imaging axis. The finite ray-origin parameter (virtual pinhole origin) is obtained during camera calibration (see Supplementary Material, Section~\textit{Imaging-System Calibration}) and then fixed for subsequent reconstructions. To reduce ambiguity, center thickness (CT) is constrained to the measured value with a narrow search window of $\pm 10\,\mu\text{m}$. Radius, edge thickness, and surface-shape terms are instead controlled through soft penalties. Measurement diversity and the resulting parameter conditioning are discussed in the Supplementary Section~\textit{Measurement Diversity and Conditioning}.

The forward ray-tracing model is based on the calibrated camera geometry as described in the Supplementary Material (Section~\textit{Imaging-System Calibration}). Since the measured quantity is pixel displacement at the sensor, the predicted displacement is obtained from the inverse correspondence of the forward ray map between the reference and lensed states. The derivations of the ray model and of the inverse mapping are given in the Supplementary Material (Sections~\textit{Forward Ray Model} and~\textit{Predicted Displacement Calculation}).

\subsubsection{Surface optimization}
The initial step in the reconstruction is to estimate the surface which best accounts for the measured distortion field. This optimization process is kept apart from the final presentation of the design parameters. The estimation of the surface is guided by the requirement of displacement consistency; it is in a subsequent fitting stage that the optimized surface is transformed into conventional optical parameters.

\textbf{Stage 1: conic parametric optimization.} The first stage uses a conic model without aspheric coefficients:
\begin{equation}\label{eq:conic}
z(r) = \frac{r^2}{R\left[1 + \sqrt{1 - (1 + k) \frac{r^2}{R^2}}\right]}
\end{equation}
where \( R \) is the radius of curvature and \( k \) is the conic constant. This stage captures the dominant geometry and corrects for position and angular misalignments of the imaging system.

\textbf{Stage 2: heightmap refinement.} After stage 1 converges, heightmaps replace the conic representation when carrying out local refinement under weak smoothness regularization and with data-dominant weighting. The single-sided spherical measurements reported here refine the front-surface heightmap and keep the preceding parametric back-surface estimate fixed. Paired front/back measurements can refine two heightmaps. In this stage, the spatial deviations from the ideal conic surface are recovered, and the dense surface estimate obtained is the one used for residual and aberration analysis.

\subsubsection{Surface fitting for final shape parameters}\label{methods:surface_fitting}
The optical shape parameters are obtained by fitting a parametric surface to the optimized heightmap. In this registration step, surface form is separated from any remaining translation and orientation, which is necessary for performing dense 3D measurements without having to use an independent datum \cite{MCBRIDE1996597,HILL2002457,elhayek2014,El-Hayek2015-qe}. Further information regarding the pose registration, the fitted parameter vector, and the optimization process is given in the Supplementary Material (Section~\textit{Reconstruction Objective and Optimization Stages}).

After this registration, the local sag follows the ISO 10110-12 asphere convention \cite{ISO10110-12-2019}:
\begin{equation}\label{eq:asphere}
z(r) = \frac{r^2}{R\left[1 + \sqrt{1 - (1 + k) \frac{r^2}{R^2}}\right]} + \sum_{i=1}^{N} A_{2i} r^{2i}
\end{equation}
In the case of spherical analysis, both \(k\) and \(A_{2i}\) are set to zero; for aspheric analysis, they are determined after the heightmap stage, up to \(A_{10}\) in the current implementation. The spherical radii from this fitting process are compared with catalog specifications. For the quantitative AL2550 benchmark, staged optimization directly fits the shared parameters $R,k,A_4,A_6,A_8$ with capture-specific lateral translation and tilt; $A_{10}$ is fixed to zero. Its CCP rotational fitting model includes $A_{10}$. The fitting model remains the same for dual-surface lenses. The larger errors, especially those occurring on the flatter back surfaces, are in agreement with the fact that the two-surface inverse problem is less well conditioned.

Residual definitions, filtering, wavefront conversion, and Zernike decomposition are described in the Supplementary Material (Section~\textit{Residual and Aberration Analysis}).

\section{Results}
We measured off-the-shelf Thorlabs lenses commonly used in optics laboratories. The set includes plano-convex and biconvex spherical lenses, and plano-aspheric lenses. Other geometries, including plano-concave, biconcave, and meniscus lenses, are compatible with the same measurement principle. The analyzed repeat counts are listed with the results.

The use of the optimized height enables the study of residual surface departures described in the Supplementary Material (Section~\textit{Residual and Aberration Analysis}). In the current arrangement, however, the analysis is limited to the low-spatial-frequency components since the imaging system, with its \(0.185\times\) telecentric magnification and \(5.5\,\mu\text{m}\) monochrome sensor, is not very sensitive to higher-spatial-frequency terms.

\subsection{Calibration results}
Four separate calibration measurements yielded an effective telecentricity error of about \(0.144^\circ\), a projective ray-origin distance of about \(4.38\,\mathrm{m}\) from the pattern plane, and an RMS displacement residual of \(0.60 \pm 0.05\,\mu\mathrm{m}\). Although these values describe the effective projective camera model, they should not be understood as a direct restatement of the manufacturer's telecentric-lens specification. Further details regarding the calibration procedure, the refractive-index measurement, the displacement maps, the repeatability statistics, and the convergence results are given in the Supplementary Material (Sections~\textit{Imaging-System Calibration} and~\textit{Experimental Refractive-Index Measurements}).

\subsection{Plano-convex lenses}
Plano-convex lenses provide the simplest example when dealing with front-surface recovery since most of the optical power is produced at a single refractive surface. For each lens model, we report the mean signed deviation from catalog radius and a two-sided 95\% confidence interval of the repeated-fit mean, computed as $\bar{x}\pm t_{0.975,n-1}s/\sqrt{n}$ with sample standard deviation $s$ and $n$ included fits. These intervals describe repeatability; they exclude calibration uncertainty and catalog tolerance. For groups with at least four fits, radius deviations outside the 1.5-interquartile-range fences are rejected before computing the mean and interval. The figure tables give included/total counts: one fit is rejected for each of LA1027-A, LA1131-A, and LA1229-A, and for the LBF254-050 front surface. Other groups include all fits. Radius fitting uses an aperture diameter of 20 mm for all models.

\begin{figure}[t]
\centering
\includegraphics[width=0.9\linewidth]{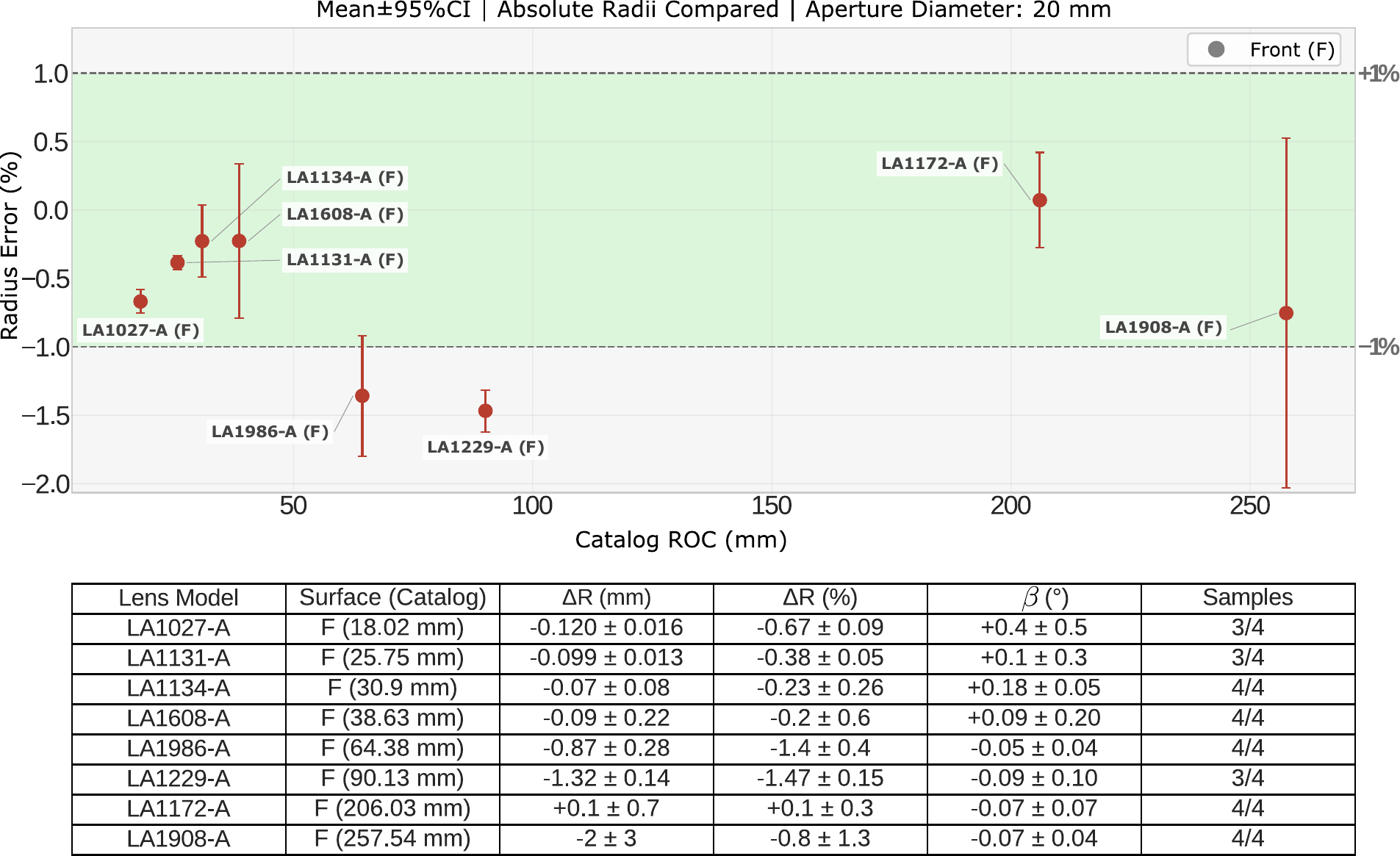}
\caption{Mean signed front-surface radius error ($\Delta R$ in \%) for plano-convex lenses, with 95\% confidence intervals from repeated measurements (aperture diameter: 20 mm). The shaded band marks the $\pm1\%$ region around nominal ROC. The accompanying table reports, for each lens, nominal catalog ROC, signed radius deviation in mm, percentage error, estimated axis tilt ($\beta$), and included/total sample count.}
\label{fig:r1_comparison}
\end{figure}

Figure~\ref{fig:r1_comparison} reports front-surface ROC estimates for eight plano-convex lenses. Six of the eight mean radius errors fall within the $\pm1\%$ band, and the mean errors span approximately $-1.5\%$ to $+0.07\%$. Most mean errors are negative, indicating a mild underestimation of front-surface radius. The lens with the largest nominal radius has the widest repeatability interval.

\subsection{Biconvex lenses}

For the biconvex lenses, we give the signed ROC errors for both the front and back surfaces of the LBF254-050 and LBF254-200, using the same 20 mm aperture criterion and the same repeated-measurement statistics.

\begin{figure}[t]
\centering
\includegraphics[width=0.9\linewidth]{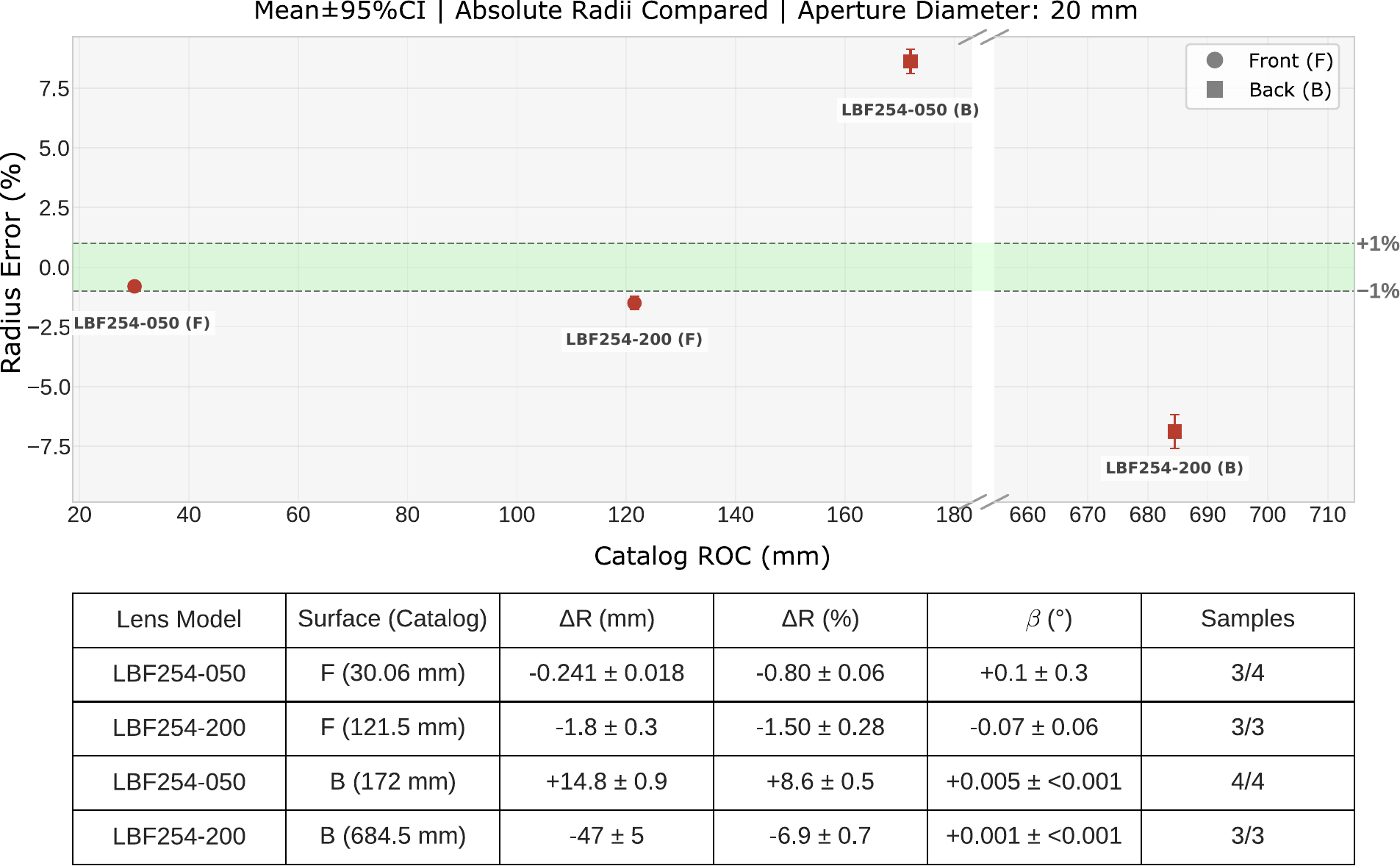}
\caption{Mean signed radius error ($\Delta R$ in \%) for front and back surfaces of two biconvex lenses (LBF254-050 and LBF254-200), with 95\% confidence intervals (aperture diameter: 20 mm). The shaded region marks the $\pm1\%$ interval. The axis break separates low/medium ROC values from the large-ROC back surface of LBF254-200. The table reports signed radius deviations in millimeters and percent, fitted axis tilt $\beta$, and included/total sample count for each surface. Angular confidence halfwidths below $0.001^\circ$ are shown as upper bounds.}
\label{fig:comprehensive_comparison_biconvex}
\end{figure}

A clear asymmetry is observed between the recoverability of the front and back surfaces in Figure~\ref{fig:comprehensive_comparison_biconvex}. The front-surface mean errors are –0.80\% and –1.5\%. The back surfaces exhibit larger systematic deviations (+8.6\% for LBF254-050 and –6.9\% for LBF254-200). The confidence intervals are narrower than the mean bias, showing that the back-surface estimates are stable but biased.

The pattern is compatible with restricted conditioning regarding back-surface recovery in two-surface lenses. The displacement data have a strong effect on the front curvature, whereas the back curvature is more affected by model mismatch and by coupling via lens thickness and refractive path. The sign reversal observed in the two back-surface cases cannot be explained by a single global offset and is compatible with a curvature-dependent bias.

\subsection{Plano aspheric lenses}

Plano-aspheric lenses require a fuller surface prescription than spherical samples: radius of curvature, conic constant, and higher-order aspheric coefficients as defined in Eq.~\eqref{eq:asphere}. We focus on the Thorlabs AL2550 lens. Because different combinations of the higher-order coefficients can produce very similar sag profiles over a finite aperture, we use physical surface agreement as the primary accuracy measure and treat individual coefficient recovery as a secondary identifiability diagnostic. The corresponding conditioning analysis is provided in the Supplementary Material.

\subsubsection{Measurement results for AL2550 aspheric lens}

For the AL2550 benchmark, direct staged optimization of a shared rotational asphere uses a fixed \(20\,\mathrm{mm}\)-diameter reconstruction aperture centered on paired fiducial-derived carrier center estimates. Lateral decenter and two-axis tilt are fitted independently for each acquisition, while in-plane rotations retain their nominal values. Center thickness is fixed to the measured value of \(6.029\,\mathrm{mm}\), and the refractive index is set to 1.5178 at 565 nm \cite{Polyanskiy2024}. The independent CCP measurement has a smaller usable field, so quantitative DRT--CCP surface comparisons are restricted to their common central \(14\,\mathrm{mm}\)-diameter region. Reconstruction stages, aperture construction, and identifiability controls are described in the Supplementary Material.

\subsubsection{Benchmark comparison with confocal chromatic profilometry}
Chromatic confocal profilometry (CCP) is used as an independent benchmark for the AL2550 rotational surface. Acquisition details, the fixed CCP validation window, and the surface-fitting procedure are described in the Supplementary Material (Section~\textit{CCP Benchmark Processing}).

Table~\ref{tab:al2550_surface_comparison} reports circular-area RMS sag differences over the common \(14\,\mathrm{mm}\)-diameter region. The rotational surface fitted to the independently acquired CCP data differs from the manufacturer prescription by \(0.09\,\mu\mathrm{m}\) RMS. The selected DRT reconstruction differs from the manufacturer surface by \(0.50\,\mu\mathrm{m}\) RMS and from the CCP rotational surface by \(0.44\,\mu\mathrm{m}\) RMS. These compare rotational surfaces, not the raw CCP point cloud, and do not constitute measurement-uncertainty bounds. Sag comparisons are used because the individual high-order coefficients are strongly correlated and poorly conditioned over the finite measurement aperture.

\begin{table}[htbp]
\centering
\caption{AL2550 rotational-surface comparison over the common central \(14\,\mathrm{mm}\)-diameter region. Values are circular-area RMS sag differences.}
\label{tab:al2550_surface_comparison}
\begin{tabular}{lc}
\toprule
Surface comparison & RMS sag difference (\(\mu\mathrm{m}\)) \\
\midrule
CCP vs. manufacturer prescription & 0.09 \\
DRT vs. manufacturer prescription & 0.50 \\
DRT vs. CCP rotational fit & 0.44 \\
\bottomrule
\end{tabular}
\end{table}

Additional optimization tests show that a small multiplicative mismatch in the measured displacement scale can be absorbed into the low-order aspheric shape. Fitting this displacement-scale factor before further shape optimization suppresses most of the resulting sag error; the conditioning and displacement-scale tests are reported in the Supplementary Material.

Controlled synthetic perturbations were also used to characterize recoverability independently of the catalog coefficient parameterization. Directional sinusoidal sag perturbations spanning 2--12 cycles per aperture were tested first with the ray-tracing inverse model and then with the complete simulated image and optical-flow pipeline. With the realistic image model, the lowest RMS sag amplitudes satisfying the recovery criteria were 5, 2, 1, and \(1\,\mu\mathrm{m}\) at 2, 4, 8, and 12 cycles per aperture, respectively. These values describe synthetic recovery of the known sinusoidal basis and are not experimental surface-resolution limits; additional basis-transfer tests are summarized in the Supplementary Material.

The residual structure obtained from the DRT and CCP reconstructions is discussed separately in the Supplementary Material (Section~\textit{Residual and Aberration Analysis}); this exploratory residual comparison is not used as the quantitative surface-accuracy metric.

\section{Conclusions}

The study deals with the measurement of the transmission properties of refractive lenses using dense mask-displacement fields. The apparatus involves a chrome wavelet-noise mask, a near-telecentric imaging system, the calibration of the non-ideal telecentricity using a finite-origin projective camera model, and differentiable ray tracing; the spherical-lens workflow proceeds from a conic surface to a heightmap and final optical parameters, and the AL2550 benchmark directly optimizes a shared rotational asphere with acquisition-specific poses.

With plano-convex spherical lenses, six of the eight mean radius errors lie within \(\pm1\%\) of the nominal catalog value. In the case of two-surface biconvex lenses, higher demands are placed on the measurements. The front surfaces are recovered with only a slight bias, but the flatter or less strongly constrained back surfaces exhibit larger systematic deviations. This result is in agreement with the nature of a transmission measurement, since multiple refracting interfaces can give rise to coupled displacement signatures.

With respect to the AL2550 plano-aspheric lens, the rotational reconstruction differs from the CCP rotational fit by $0.44\,\mu\mathrm{m}$ RMS over the common validation aperture. Individual high-order polynomial coefficients remain poorly conditioned and are therefore not interpreted independently as accuracy measures. Supplementary optimization and displacement-scale tests show that a small multiplicative displacement error can be absorbed by the low-order aspheric shape, which supports direct validation of the recovered sag against an independent surface measurement. There are still some limitations. The calibration model currently does not take field-dependent imaging distortion into account. The conditioning of dual-surface lenses needs to be improved, particularly when the back surfaces are flatter. The aberration terms at high spatial frequencies are limited by the existing imaging chain. Overcoming these limitations will demand field-dependent camera calibration, higher-resolution imaging, and independent verification of the recovered residual surface. Since the method makes use of non-coherent illumination and a reusable reference image, each lens measurement only necessitates the acquisition of the distorted pattern image. The working distance is determined by the telecentric imaging system and does not involve the need for a near-contact probe.

\section*{Funding}
King Abdullah University of Science and Technology (KAUST) (Individual Baseline Funding).

\section*{Acknowledgments}
The authors thank Qiang Fu and Hadi Amata for their help with the binary mask fabrication.

\section*{Disclosures}
The authors declare no conflicts of interest.

\section*{Data availability}
Data underlying the results presented in this paper are not publicly available at this time but may be obtained from the authors upon reasonable request.

\section*{Supplemental document}
The supplementary material follows the main manuscript in this arXiv submission.

\bibliography{references}

\begin{thebibliography}{10}
\newcommand{\enquote}[1]{``#1''}

\bibitem{elhayek2014}
N.~El-Hayek, H.~Nouira, N.~Anwer, \emph{et~al.}, \enquote{A new method for
  aspherical surface fitting with large-volume datasets,}
  {\protect\JournalTitle{Precision Engineering}} \textbf{38}, 935--947 (2014).

\bibitem{ye2021}
J.~Ye, Z.~Niu, X.~Zhang, \emph{et~al.}, \enquote{Simultaneous measurement of
  double surfaces of transparent lenses with phase measuring deflectometry,}
  {\protect\JournalTitle{Optics and Lasers in Engineering}} \textbf{137},
  106356 (2021).

\bibitem{hao2023}
Q.~Hao, Y.~Liu, Y.~Hu, and X.~Tao, \enquote{Measurement techniques for aspheric
  surface parameters,} {\protect\JournalTitle{Light: Advanced Manufacturing}}
  \textbf{4} (2023).

\bibitem{mishra2024}
V.~Mishra, D.~R. Burada, K.~K. Pant, and G.~S. Khan, \enquote{Advances in
  ultra-precision manufacturing and metrology for freeform optics,} in
  \emph{Advances in Optical Manufacturing,}  (Elsevier, 2024).

\bibitem{Rolland:21}
J.~P. Rolland, M.~A. Davies, T.~J. Suleski, \emph{et~al.}, \enquote{Freeform
  optics for imaging,} {\protect\JournalTitle{Optica}} \textbf{8}, 161--176
  (2021).

\bibitem{Petz2009-yd}
M.~Petz, M.~Fischer, and R.~Tutsch, \enquote{Three-dimensional shape
  measurement of aspheric refractive optics by pattern transmission
  photogrammetry,} in \emph{{Three-Dimensional} Imaging Metrology,}  J.~A.
  Beraldin, G.~S. Cheok, M.~McCarthy, and U.~Neuschaefer-Rube, eds. (SPIE,
  2009).

\bibitem{beutler2016}
A.~Beutler, \enquote{Metrology for the production process of aspheric lenses,}
  {\protect\JournalTitle{Advanced Optical Technologies}} \textbf{5}, 211--228
  (2016).

\bibitem{ramirez2021}
A.~H. Ramirez-Andrade, S.~Shadalou, D.~Gurganus, \emph{et~al.}, \enquote{Vision
  ray metrology for freeform optics,} {\protect\JournalTitle{Optics Express}}
  \textbf{29}, 43480--43501 (2021).

\bibitem{miao2026}
L.~Miao, K.~Liu, Z.~Wang, \emph{et~al.}, \enquote{Towards industrial in-field
  optical metrology: A wavefront aberrations retrieval technique through
  single-frame interferogram,} {\protect\JournalTitle{Measurement}}
  \textbf{262}, 120049 (2026).

\bibitem{wyant1974}
J.~C. Wyant and P.~K. O'Neill, \enquote{Computer generated hologram; null lens
  test of aspheric wavefronts,} {\protect\JournalTitle{Applied Optics}}
  \textbf{13}, 2762--2765 (1974).

\bibitem{diaz2000}
R.~Díaz-Uribe and M.~Campos-García, \enquote{Null-screen testing of fast
  convex aspheric surfaces,} {\protect\JournalTitle{Applied Optics}}
  \textbf{39}, 2670--2677 (2000).

\bibitem{knauer2004}
M.~C. Knauer, J.~Kaminski, and G.~Hausler, \enquote{Phase measuring
  deflectometry: a new approach to measure specular free-form surfaces,}
  {\protect\JournalTitle{Proc. SPIE}} \textbf{5457}, 366--376 (2004).

\bibitem{huerta2021}
O.~Huerta-Carranza, M.~Avendaño-Alejo, and R.~Díaz-Uribe, \enquote{Null
  screens to evaluate the shape of freeform surfaces: progressive addition
  lenses,} {\protect\JournalTitle{Optics Express}} \textbf{29}, 27921--27937
  (2021).

\bibitem{wang2021}
C.~Wang, N.~Chen, and W.~Heidrich, \enquote{Towards self-calibrated lens
  metrology by differentiable refractive deflectometry,}
  {\protect\JournalTitle{Optics Express}} \textbf{29}, 30284--30295 (2021).

\bibitem{burke2023}
J.~Burke, A.~Pak, S.~H{\"o}fer, \emph{et~al.}, \enquote{Deflectometry for
  specular surfaces: an overview,} {\protect\JournalTitle{Adv. Opt. Technol.}}
  \textbf{12} (2023).

\bibitem{Raffel2015}
M.~Raffel, \enquote{Background-oriented schlieren ({BOS}) techniques,}
  {\protect\JournalTitle{Experiments in Fluids}} \textbf{56} (2015).

\bibitem{Moisy2009}
F.~Moisy, M.~Rabaud, and K.~Salsac, \enquote{A synthetic schlieren method for
  the measurement of the topography of a liquid interface,}
  {\protect\JournalTitle{Experiments in Fluids}} \textbf{46}, 1021–1036
  (2009).

\bibitem{Cakir2022}
B.~O. Cakir, S.~Lavagnoli, B.~H. Saracoglu, and C.~Fureby, \enquote{Assessment
  and application of optical flow in background-oriented schlieren for
  compressible flows,} {\protect\JournalTitle{Experiments in Fluids}}
  \textbf{64} (2022).

\bibitem{Ota2015}
M.~Ota, F.~Leopold, R.~Noda, and K.~Maeno, \enquote{Improvement in spatial
  resolution of background-oriented schlieren technique by introducing a
  telecentric optical system and its application to supersonic flow,}
  {\protect\JournalTitle{Experiments in Fluids}} \textbf{56} (2015).

\bibitem{Fischer2014}
M.~Fischer, M.~Petz, and R.~Tutsch, \enquote{Model-based deflectometric
  measurement of transparent objects,} in \emph{Fringe 2013,}  W.~Osten, ed.
  (Springer Berlin Heidelberg, Berlin, Heidelberg, 2014), pp. 573--576.

\bibitem{huerta2024}
O.~Huerta-Carranza, F.~S. Granados-Agust{\'\i}n, A.~Santiago-Alvarado,
  \emph{et~al.}, \enquote{Evaluation of the aspherical surface of a
  plano-convex lens by refraction using an {LCD},} {\protect\JournalTitle{Rev.
  Sci. Instrum.}} \textbf{95} (2024).

\bibitem{ramirez2024}
A.~H. Ramirez-Andrade and K.~Falaggis, \enquote{Height reconstructions from
  geometric wavefronts using vision ray metrology,}
  {\protect\JournalTitle{Applied Optics}} \textbf{63}, 8630 (2024).

\bibitem{Xu:2019}
D.~Xu, R.~Chaudhuri, and J.~P. Rolland, \enquote{Telecentric broadband
  objective lenses for optical coherence tomography ({OCT}) in the context of
  low uncertainty metrology of freeform optical components: from design to
  testing for wavefront and telecentricity,} {\protect\JournalTitle{Opt.
  Express}} \textbf{27}, 6184--6200 (2019).

\bibitem{MAESTRO-WATSON2017}
D.~Maestro-Watson, A.~Izaguirre, and N.~Arana-Arexolaleiba, \enquote{{LCD}
  screen calibration for deflectometric systems considering a single layer
  refraction model,} in \emph{2017 IEEE International Workshop of Electronics,
  Control, Measurement, Signals and their Application to Mechatronics (ECMSM),}
   (2017), pp. 1--6.

\bibitem{Berto2017}
P.~Berto, H.~Rigneault, and M.~Guillon, \enquote{Wavefront sensing with a thin
  diffuser,} {\protect\JournalTitle{Optics Letters}} \textbf{42}, 5117 (2017).

\bibitem{Wang2017}
C.~Wang, X.~Dun, Q.~Fu, and W.~Heidrich, \enquote{Ultra-high resolution coded
  wavefront sensor,} {\protect\JournalTitle{Optics Express}} \textbf{25},
  13736--13746 (2017).

\bibitem{Wang2019}
C.~Wang, Q.~Fu, X.~Dun, and W.~Heidrich, \enquote{Quantitative phase and
  intensity microscopy using snapshot white light wavefront sensing,}
  {\protect\JournalTitle{Scientific Reports}} \textbf{9}, 13795 (2019).

\bibitem{Cook2005}
R.~L. Cook and T.~DeRose, \enquote{Wavelet noise,} {\protect\JournalTitle{ACM
  Trans. Graph.}} \textbf{24}, 803–811 (2005).

\bibitem{RN592}
B.~Atcheson, W.~Heidrich, and I.~Ihrke, \enquote{An evaluation of optical flow
  algorithms for background oriented schlieren imaging,}
  {\protect\JournalTitle{Experiments in Fluids}} \textbf{46}, 467--476 (2009).

\bibitem{Farneback2003}
G.~Farnebäck, \enquote{Two-frame motion estimation based on polynomial
  expansion,} in \emph{Image Analysis,}  vol. 2749 of \emph{Lecture Notes in
  Computer Science} J.~Bigun and T.~Gustavsson, eds. (Springer, 2003), pp.
  363--370.

\bibitem{WANG2024102029}
Y.~Wang, W.~Wang, Y.~Li, \emph{et~al.}, \enquote{Research on traditional and
  deep learning strategies based on optical flow estimation - a review,}
  {\protect\JournalTitle{Journal of King Saud University - Computer and
  Information Sciences}} \textbf{36}, 102029 (2024).

\bibitem{FarnebackOpenCV}
{OpenCV}, \enquote{Optical flow: {calcOpticalFlowFarneback},}
  \url{https://docs.opencv.org/3.4.20/dc/d6b/group__video__track.html}.
  Accessed October 8, 2026.

\bibitem{MCBRIDE1996597}
J.~McBride and K.~Cross, \enquote{The measurement and analysis of the
  three-dimensional form of curved surfaces,}
  {\protect\JournalTitle{International Journal of Machine Tools and
  Manufacture}} \textbf{36}, 597--610 (1996).

\bibitem{HILL2002457}
M.~Hill, M.~Jung, and J.~McBride, \enquote{Separation of form from orientation
  in {3D} measurements of aspheric surfaces with no datum,}
  {\protect\JournalTitle{International Journal of Machine Tools and
  Manufacture}} \textbf{42}, 457--466 (2002).

\bibitem{El-Hayek2015-qe}
N.~El-Hayek, N.~Anwer, H.~Nouira, \emph{et~al.}, \enquote{{3D} measurement and
  characterization of ultra-precision aspheric surfaces,}
  {\protect\JournalTitle{Procedia CIRP}} \textbf{27}, 41--46 (2015).

\bibitem{ISO10110-12-2019}
{International Organization for Standardization}, \enquote{{ISO} 10110-12:2019
  optics and photonics — preparation of drawings for optical elements and
  systems — part 12: Aspheric surfaces,} Standard ISO 10110-12:2019,
  International Organization for Standardization, Geneva (2019).

\bibitem{Polyanskiy2024}
M.~N. Polyanskiy, \enquote{Refractiveindex.info database of optical constants,}
  {\protect\JournalTitle{Scientific Data}} \textbf{11} (2024).

\end{thebibliography}


\begin{thebibliography}{10}
\newcommand{\enquote}[1]{``#1''}

\bibitem{Cook2005}
R.~L. Cook and T.~DeRose, \enquote{Wavelet noise,} {\protect\JournalTitle{ACM
  Trans. Graph.}} \textbf{24}, 803–811 (2005).

\bibitem{Farneback2003}
G.~Farnebäck, \enquote{Two-frame motion estimation based on polynomial
  expansion,} in \emph{Image Analysis,}  vol. 2749 of \emph{Lecture Notes in
  Computer Science} J.~Bigun and T.~Gustavsson, eds. (Springer, 2003), pp.
  363--370.

\bibitem{WANG2024102029}
Y.~Wang, W.~Wang, Y.~Li, \emph{et~al.}, \enquote{Research on traditional and
  deep learning strategies based on optical flow estimation - a review,}
  {\protect\JournalTitle{Journal of King Saud University - Computer and
  Information Sciences}} \textbf{36}, 102029 (2024).

\bibitem{Gonzalez2018section}
R.~C. Gonzalez and R.~E. Woods, \emph{Digital Image Processing} (Pearson,
  2018), 4th ed.

\bibitem{Gross2005}
H.~Gross, \enquote{Raytracing,} in \emph{Handbook of Optical Systems,}  vol.~1
  H.~Gross, ed. (Wiley, 2005), chap.~5, pp. 173--228.

\bibitem{MCBRIDE1996597}
J.~McBride and K.~Cross, \enquote{The measurement and analysis of the
  three-dimensional form of curved surfaces,}
  {\protect\JournalTitle{International Journal of Machine Tools and
  Manufacture}} \textbf{36}, 597--610 (1996).

\bibitem{HILL2002457}
M.~Hill, M.~Jung, and J.~McBride, \enquote{Separation of form from orientation
  in {3D} measurements of aspheric surfaces with no datum,}
  {\protect\JournalTitle{International Journal of Machine Tools and
  Manufacture}} \textbf{42}, 457--466 (2002).

\bibitem{elhayek2014}
N.~El-Hayek, H.~Nouira, N.~Anwer, \emph{et~al.}, \enquote{A new method for
  aspherical surface fitting with large-volume datasets,}
  {\protect\JournalTitle{Precision Engineering}} \textbf{38}, 935--947 (2014).

\bibitem{El-Hayek2015-qe}
N.~El-Hayek, N.~Anwer, H.~Nouira, \emph{et~al.}, \enquote{{3D} measurement and
  characterization of ultra-precision aspheric surfaces,}
  {\protect\JournalTitle{Procedia CIRP}} \textbf{27}, 41--46 (2015).

\bibitem{ISO10110-12-2019}
{International Organization for Standardization}, \enquote{{ISO} 10110-12:2019
  optics and photonics — preparation of drawings for optical elements and
  systems — part 12: Aspheric surfaces,} Standard ISO 10110-12:2019,
  International Organization for Standardization, Geneva (2019).

\bibitem{kingma2017adam}
D.~P. Kingma and J.~Ba, \enquote{{Adam}: A method for stochastic optimization,}
   (2017).

\bibitem{Liu1989}
D.~C. Liu and J.~Nocedal, \enquote{On the limited memory {BFGS} method for
  large scale optimization,} {\protect\JournalTitle{Mathematical Programming}}
  \textbf{45}, 503--528 (1989).

\bibitem{ISO2025}
{International Organization for Standardization}, \enquote{Geometrical product
  specifications ({GPS}) -- filtration -- part 21: Linear profile filters:
  Gaussian filters,} Standard ISO 16610-21:2025, International Organization for
  Standardization, Geneva, Switzerland (2025).

\bibitem{Mahajan2007Zernike}
V.~N. Mahajan, \enquote{Zernike polynomial and wavefront fitting,} in
  \emph{Optical Shop Testing,}  D.~Malacara, ed. (Wiley, 2007), chap.~13, pp.
  498--546.

\bibitem{Rimmer:75}
M.~P. Rimmer and J.~C. Wyant, \enquote{Evaluation of large aberrations using a
  lateral-shear interferometer having variable shear,}
  {\protect\JournalTitle{Appl. Opt.}} \textbf{14}, 142--150 (1975).

\bibitem{Svechnikov:15}
M.~V. Svechnikov, N.~I. Chkhalo, M.~N. Toropov, and N.~N. Salashchenko,
  \enquote{Resolving capacity of the circular zernike polynomials,}
  {\protect\JournalTitle{Opt. Express}} \textbf{23}, 14677--14694 (2015).

\bibitem{ramirez2024}
A.~H. Ramirez-Andrade and K.~Falaggis, \enquote{Height reconstructions from
  geometric wavefronts using vision ray metrology,}
  {\protect\JournalTitle{Applied Optics}} \textbf{63}, 8630 (2024).

\bibitem{Zhang2000}
Z.~Zhang, \enquote{A flexible new technique for camera calibration,}
  {\protect\JournalTitle{IEEE Transactions on Pattern Analysis and Machine
  Intelligence}} \textbf{22}, 1330--1334 (2000).

\bibitem{LI2013538}
D.~Li and J.~Tian, \enquote{An accurate calibration method for a camera with
  telecentric lenses,} {\protect\JournalTitle{Optics and Lasers in
  Engineering}} \textbf{51}, 538--541 (2013).

\bibitem{Duan:22}
Y.~Duan, P.~Hu, H.~Deng, \emph{et~al.}, \enquote{Flexible and simple
  telecentric camera calibration method,} {\protect\JournalTitle{Appl. Opt.}}
  \textbf{61}, 9625--9633 (2022).

\bibitem{Hartley2003}
R.~Hartley and A.~Zisserman, \emph{Multiple View Geometry in Computer Vision}
  (Cambridge University Press, 2004), 2nd ed.

\bibitem{ZONG2026109854}
Y.~Zong, W.~Li, X.~Wang, \emph{et~al.}, \enquote{Accurate geometric modeling
  and calibration of bi-telecentric imaging systems for precision optical
  measurement,} {\protect\JournalTitle{Optics and Lasers in Engineering}}
  \textbf{204}, 109854 (2026).

\bibitem{Xu:2019}
D.~Xu, R.~Chaudhuri, and J.~P. Rolland, \enquote{Telecentric broadband
  objective lenses for optical coherence tomography ({OCT}) in the context of
  low uncertainty metrology of freeform optical components: from design to
  testing for wavefront and telecentricity,} {\protect\JournalTitle{Opt.
  Express}} \textbf{27}, 6184--6200 (2019).

\bibitem{Triggs2000BundleAdjustment}
B.~Triggs, P.~F. McLauchlan, R.~I. Hartley, and A.~W. Fitzgibbon,
  \enquote{Bundle adjustment --- a modern synthesis,} in \emph{Vision
  Algorithms: Theory and Practice,}  B.~Triggs, A.~Zisserman, and R.~Szeliski,
  eds. (Springer Berlin Heidelberg, Berlin, Heidelberg, 2000), pp. 298--372.

\bibitem{Penrose1955GeneralizedInverse}
R.~Penrose, \enquote{A generalized inverse for matrices,}
  {\protect\JournalTitle{Mathematical Proceedings of the Cambridge
  Philosophical Society}} \textbf{51}, 406--413 (1955).

\bibitem{GolubPereyra1973}
G.~H. Golub and V.~Pereyra, \enquote{The differentiation of pseudo-inverses and
  nonlinear least squares problems whose variables separate,}
  {\protect\JournalTitle{SIAM Journal on Numerical Analysis}} \textbf{10},
  413--432 (1973).

\bibitem{nanovea}
B.~Mell, \enquote{Aspheric lens dimension using {3D} profilometry,} Tech. rep.,
  Nanovea (2010).

\bibitem{Blateyron2011}
F.~Blateyron, \enquote{Chromatic confocal microscopy,} in \emph{Optical
  Measurement of Surface Topography,}  R.~Leach, ed. (Springer Berlin
  Heidelberg, Berlin, Heidelberg, 2011), pp. 71--106.

\bibitem{hao2023}
Q.~Hao, Y.~Liu, Y.~Hu, and X.~Tao, \enquote{Measurement techniques for aspheric
  surface parameters,} {\protect\JournalTitle{Light: Advanced Manufacturing}}
  \textbf{4} (2023).

\end{thebibliography}

\end{document}


\title{Supplement: Non-contact lens metrology from optical flow and differentiable ray tracing}

\author{Hector A. Jimenez-Romero,\authormark{1} and Wolfgang Heidrich\authormark{1,*}}

\address{\authormark{1}King Abdullah University of Science and Technology (KAUST), Thuwal 23955-6900, Saudi Arabia}
\email{\authormark{*}wolfgang.heidrich@kaust.edu.sa}

\section{Pattern Design and Capture Fidelity}\label{sec:supp_pattern_fidelity}
\subsection{Binary mask design}

A chrome-etched binary wavelet-noise mask provides high contrast and multiscale features suitable for displacement estimation and optical-flow processing \cite{Cook2005,Farneback2003,WANG2024102029}. The fabricated chrome feature size is $1.975\,\mu\text{m}$. This way, the effective sampling of the camera converts the binary pattern into a grayscale target, with enough contrast for sub-pixel-level tracking. 

Given telecentric magnification $M\approx0.185\times$, the effective object-space sampling is
\begin{equation}
\delta=\frac{p_{\mathrm{cam}}}{M},
\end{equation}
where $p_{\mathrm{cam}}$ is sensor pixel pitch. With a $5.5\,\mu\text{m}$ monochrome sensor, $\delta\approx29.7\,\mu\text{m}$. The mask sampling was chosen so that one camera pixel integrates approximately one $15$-feature binary block, since $15\times1.975\,\mu\text{m}\approx29.6\,\mu\text{m}$.

\subsection{Pattern-generation workflow}
The mask design starts from a grayscale wavelet-noise field generated by summing band-pass noise components over multiple spatial scales. This grayscale target is converted to a binary fabrication raster by positional dithering, after which fiducial markers and alignment guides are overlaid. The result contains multi-scale texture for optical-flow tracking.

\begin{figure}[htbp]
\centering
\includegraphics[width=0.5\linewidth]{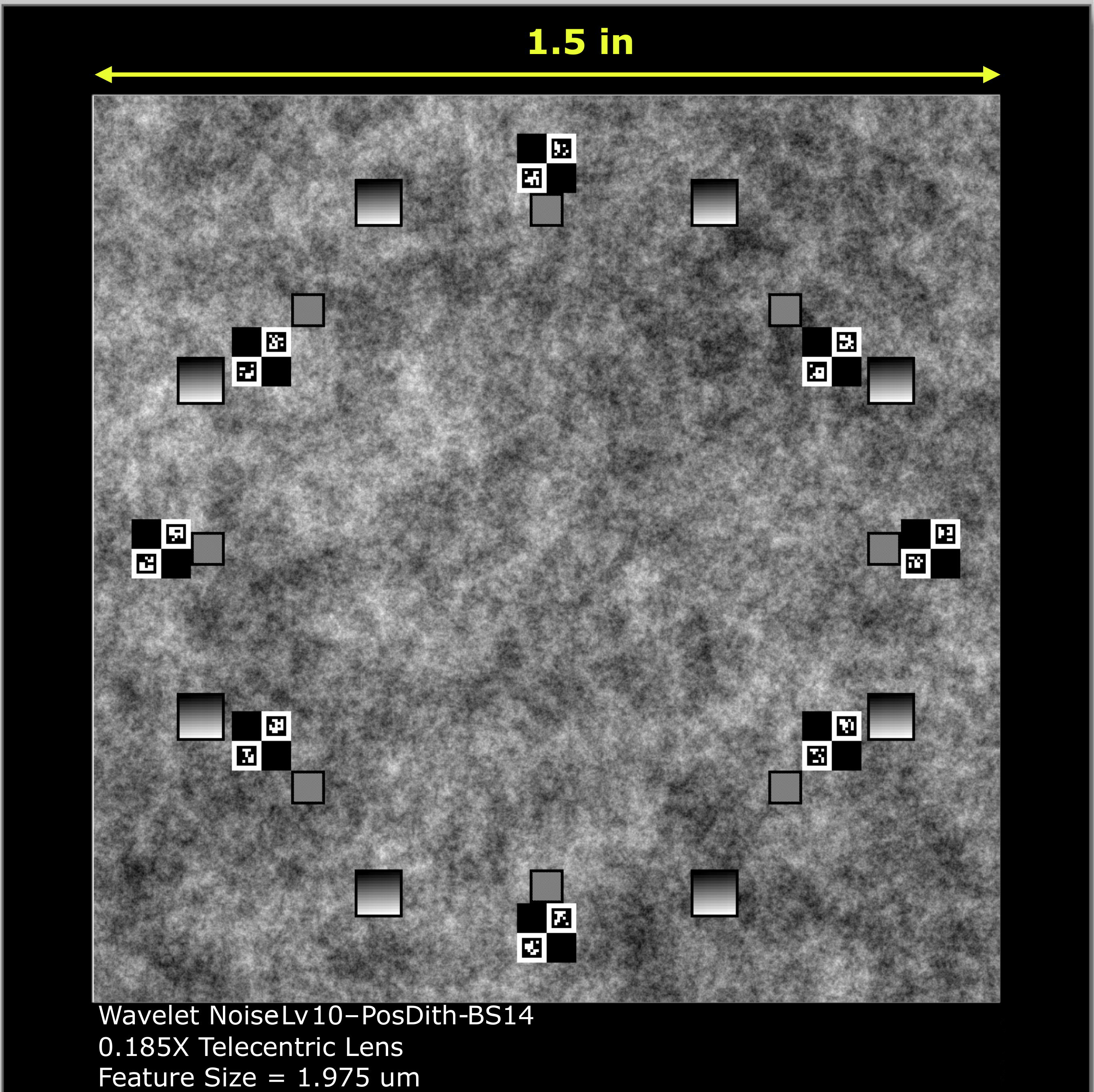}
\caption{Layout representation of the fabricated wavelet-noise mask. The central square-textured region provides the dense tracking pattern used for optical flow estimation. Fiducial markers and alignment guides are included for registration and magnification corrections. The figure shows a layout rendering; the full-resolution binary raster is not reproduced.}
\label{fig:supp_mask_layout}
\end{figure}

The final fabrication raster is $19260\times19260$ pixels. Figure \ref{fig:supp_mask_layout} shows the layout representation of the fabricated mask. With a $1.975\,\mu\text{m}$ fabrication feature size, the active design spans approximately $38.0\,\mathrm{mm}$, around $1.5\,\text{in}$. The effective sensor-simulation block is $15$ fabrication pixels, or $29.6\,\mu\text{m}$, which closely matches the object-space camera sampling. The binary chrome mask was fabricated by direct laser writing on a Heidelberg Instruments uPG101 system. Table \ref{tab:supp_mask_generation} summarizes the relevant mask-design parameters.

\begin{table}[htbp]
\centering
\caption{Relevant mask-design parameters for the wavelet-noise target used in the reported measurements.}
\label{tab:supp_mask_generation}
\begin{tabular}{ll}
\toprule
Quantity & Value \\
\midrule
Fabrication feature size & $1.975\,\mu\text{m}$ \\
Final raster size & $19260\times19260$ pixels \\
Approximate physical span & $38.0\,\mathrm{mm}$ \\
Effective sensor-simulation block & $15$ fabrication pixels \\
Object-space block size & $29.6\,\mu\text{m}$ \\
Fabrication system & Heidelberg Instruments uPG101 \\
\bottomrule
\end{tabular}
\end{table}

\subsection{Fabrication and image-domain validation}
Pattern suitability is assessed from photometric agreement between the designed and captured reference pattern, residual maps between reference and lensed captures, and histogram/PSD checks over the spatial-frequency band used by optical flow.

\begin{figure}[htbp]
\centering
\includegraphics[width=1.0\linewidth]{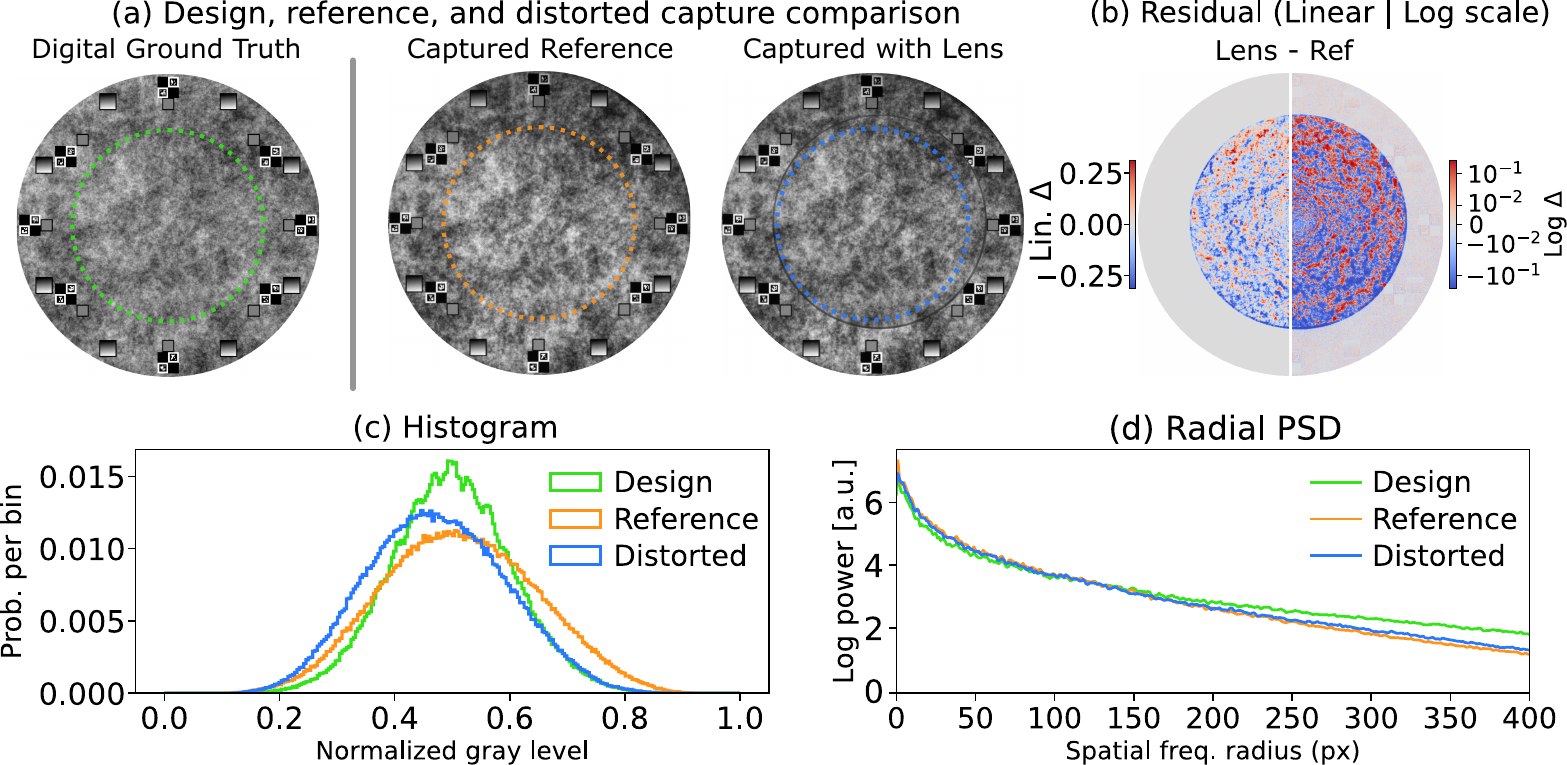}
\caption{Comparison of wavelet noise pattern fidelity and distortion residuals. (a) Normalized grayscale images of the design, measured reference, and distorted by the lens capture. Dashed circles denote the $400$\,px analysis radius. (b) Spatial map of the distortion residual (Lens distorted $-$ Reference) using a split-scale visualization: the left hemisphere uses a linear scale to display absolute error magnitudes, and the right hemisphere uses a symmetrical logarithmic scale to highlight fine-grained residuals in the non-lens region. Independent colorbars for each scale are provided on the margins. (c) Probability histograms of the normalized gray levels for the design and captured patterns within the analysis region. (d) Radially averaged power spectral density (PSD) comparing the spatial frequency distributions of the design and captured patterns. Distorted capture corresponds to the Thorlabs aspherical lens AL2550.}

\label{fig:pattern_and_typical}
\end{figure}

Figure~\ref{fig:pattern_and_typical} compares the digital design, the reference capture, and a distorted capture through the lens. The dashed circles mark the analysis region. The residual image (distorted minus reference) contains lens-induced deviations over several spatial scales, and the split linear/symmetric-log display makes both global variations and fine artifacts visible. The gray-level histograms stay similar apart from the darker distorted image, which is caused by attenuation through the lens. The radially averaged PSD shows that the spatial frequencies needed for optical flow remain present after fabrication and imaging.

\section{Image processing}
Each measurement uses a reference pattern image and a lens-distorted pattern image, stored as lossless TIFFs. Frame averaging reduces capture noise; single-shot acquisition is possible when the signal-to-noise ratio is sufficient. An edge detector locates the lens boundary, estimates the geometric center, and builds a circular soft mask for the region used in displacement-field retrieval. The lens image is darker than the reference image, so histogram matching is applied to the distorted capture before displacement-field retrieval \cite[Sec.~3.3]{Gonzalez2018section}. For the measurements reported here, the displacement field is initialized with the physical radial model and then refined by direct image-consistency optimization, as described in Section~\textit{Initial State Estimation of the Displacement Field}. Farnebäck optical flow was also tested as an alternative initialization. Because the measured displacement is produced by refraction through the lens, we refer to the resulting field as the distortion field.

\subsection{Computational implementation}
The reconstruction code is implemented in PyTorch. Distortion-field retrieval, differentiable ray-tracing optimization, and final surface fitting are separate modules. The surface-fitting module can also be applied to external heightmaps such as profilometric measurements. Representative single-pair spherical-lens processing runs on $950\times950$ grids required approximately 5--11 min on one NVIDIA RTX A4500 GPU with 20 GB memory. These timings include displacement-field retrieval, staged surface optimization with heightmap refinement, and final parametric fitting. Within these runs, displacement-field computation alone required approximately 5--6 s per pair, including radial initialization and dense image-consistency refinement. Diagnostic plotting and artifact export brought the displacement module's total runtime to approximately 20--23 s.

On a server with two Intel Xeon Gold 6330 CPUs, the spherical reconstruction pipelines required approximately 5--15 min per capture pair on one NVIDIA A100 GPU with 80 GB memory, using precomputed displacement fields on $950\times950$ grids. These wall times include preprocessing, staged optimization, final parametric fitting, and output generation. An individual exploratory AL2550 heightmap reconstruction required approximately 8.5 min on the same hardware. These timings exclude displacement-field retrieval and subsequent independent validation.

The ten-acquisition AL2550 rotational benchmark used one A100 GPU on the same server, running PyTorch 2.6 with CUDA 12.4 and double-precision reconstruction. Retrieval of the ten input displacement fields required approximately 20--50 s per field, totaling about 6 min of computation. The joint fit to all ten acquisitions required approximately 4.5 h of accumulated optimization time across five stages totaling 11,900 iterations on $950\times950$ grids. This timing excludes displacement-field retrieval, output generation, and subsequent validation. Runtime depends on the number of acquisitions, grid size, surface representation, iteration budgets, and stopping criteria.

\section{Initial State Estimation of the Displacement Field}\label{sec:supp_initial_state}
The displacement field is initialized with a compact radial model fitted to sparse image correspondences. This model seeds the subsequent dense image-consistency refinement; the refined displacement field is then passed to the differentiable ray-tracing surface reconstruction.
\subsection{Physical model for axisymmetric lenses with a conic refractive surface}
Consider a lens with a planar side and a conic refractive surface. Let $s(r)=dz/dr$ be the radial surface slope. Explicitly, for the conic surface without aspheric correction, the slope is
\begin{equation}
 s(r)=\frac{r}{R\sqrt{1-(1+k)r^2/R^2}}.
\end{equation}
An axially incident ray sees a local surface tilt
\begin{equation}
 \theta_s=\arctan s(r).
\end{equation}
For a single refracting interface from index $n_1$ to $n_2$, Snell's law gives the transmitted angle with respect to the surface normal as
\begin{equation}
 \theta_t=\arcsin\!\left(\frac{n_1}{n_2}\sin\theta_s\right).
\end{equation}
The ray deviation from the optical axis is $\theta_s-\theta_t$, and propagation over an effective distance $L$ gives
\begin{equation}
 u_r(r)=L\tan\!\left[\theta_s-\arcsin\!\left(\frac{n_1}{n_2}\sin\theta_s\right)\right].
\end{equation}

For initialization, the compact non-paraxial approximation is used
\begin{equation}
 u_r(r)\approx \alpha\tan\!\left[\beta\arctan s(r)\right],
 \label{eq:supp_initial_tangent_model}
\end{equation}
where $\alpha$ absorbs effective propagation distance and unit conversion. The factor $\beta$ is set from the nominal refractive-index contrast and represents the leading angular deviation from Snell refraction. Keeping the outer tangent and the slope angle $\arctan s(r)$ preserves the finite-angle behavior needed near the aperture edge. In the small-slope limit, Eq.~\eqref{eq:supp_initial_tangent_model} reduces to
\begin{equation}
 u_r(r)\approx \alpha\beta s(r),
\end{equation}
which is the known paraxial relationship between displacement and surface slope. The parameter $\beta$ is fixed from nominal refractive indices for the lens material and surrounding medium. The parameters $\alpha$, $R$, and $k$ are then fitted to the observed distortion field by least squares. This estimate provides a rapid geometric identification of the sample and can be useful during manufacturing inspection. It is nevertheless a reduced model: it assumes rotational symmetry, compresses propagation and refraction into two scalar factors, and does not account for the calibrated camera geometry or coupled multi-surface effects. We therefore use it primarily to initialize the dense displacement-field refinement. The optimized displacement field is then used by the differentiable ray-tracing reconstruction.

\begin{figure}[htbp]
\centering
\includegraphics[width=0.9\linewidth]{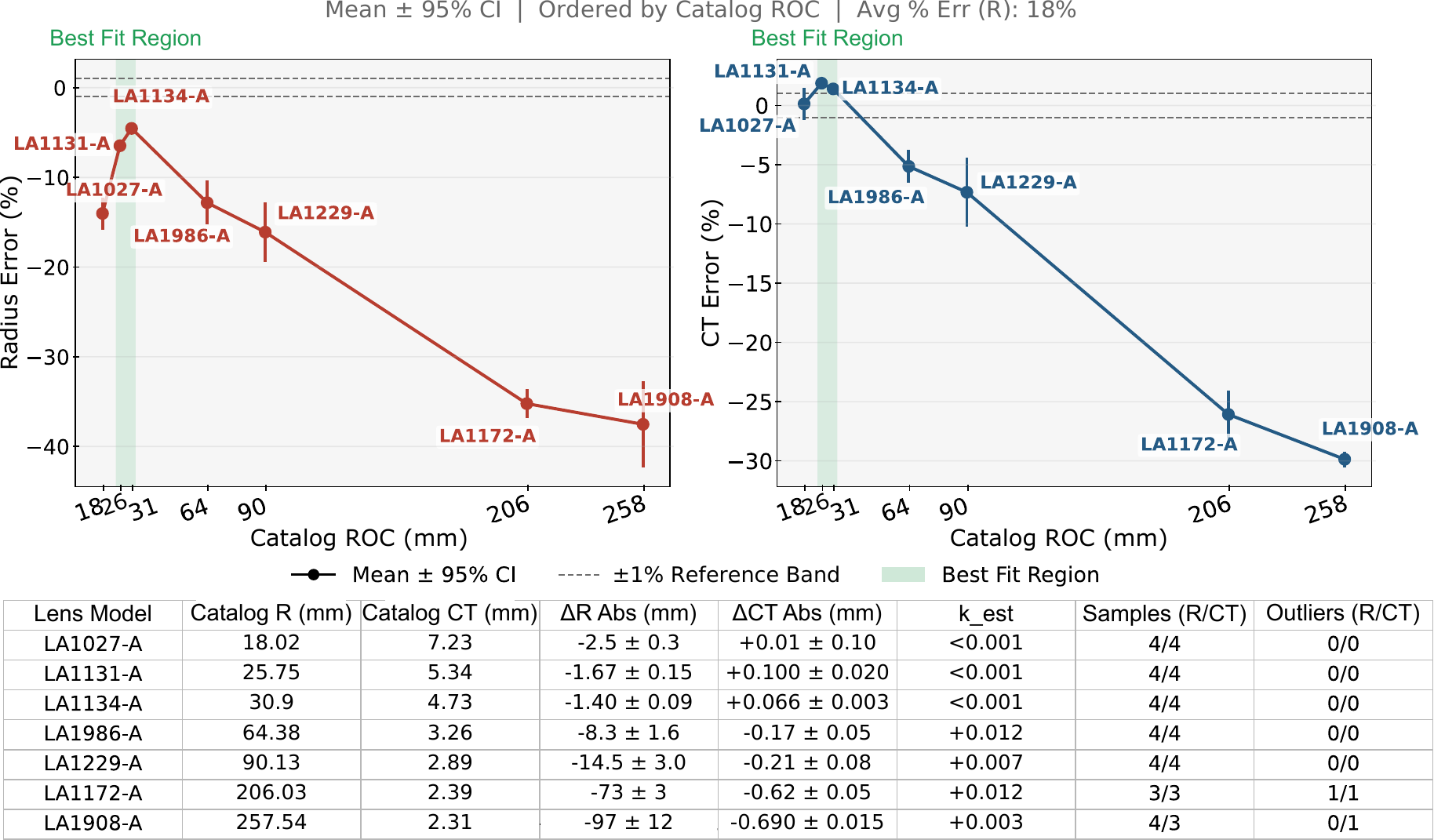}
\caption{Radial-model geometric estimates for plano-convex lenses, obtained from the refined displacement fields before ray-tracing surface reconstruction. Left: signed radius-of-curvature error. Right: signed center-thickness (CT) error. Points show repeated-fit means with two-sided 95\% confidence intervals, calculated as $\bar{x}\pm t_{0.975,n-1}s/\sqrt{n}$, where $n$ is the number of included fits, $s$ is the sample standard deviation, and $t_{0.975,n-1}$ is the 97.5th percentile of the $t$ distribution with $n-1$ degrees of freedom. These intervals describe repeatability, excluding calibration uncertainty and catalog tolerance. The shaded region marks the range with the smallest radial-model bias. The table lists the nominal geometry, signed deviations in millimeters, fitted conic constant, included radius/CT counts, and rejected outliers. Radius and CT are filtered independently using 1.5-interquartile-range fences for groups with at least four fits. Conic constants with magnitude below $0.001$ are shown as bounds.}
\label{fig:lens_init_estimation}
\end{figure}

Figure \ref{fig:lens_init_estimation} summarizes coarse geometric estimates obtained by refitting the radial model to the refined displacement fields for selected plano-convex lenses. For lenses with nominal radius of curvature below $100\,\mathrm{mm}$, the radius estimate remains within about $20\%$ of the nominal value, and the center-thickness (CT) estimate remains within about $10\%$. Flatter lenses generate weaker displacement fields and show larger radial-model bias.

\subsection{Differentiable optimization refinement}
After radial initialization, the displacement field is refined by direct image-consistency optimization. The distorted image is histogram-matched to the reference image, and the dense displacement map is optimized with a PyTorch implementation on a CUDA device. The merit function relies on three terms:

\begin{align}
\mathcal{L}_{\text{data}} &= \frac{\sum_i (I_{\text{warp}} - I_{\text{ref}})^2 \cdot m_i}{\sum_i m_i}, \\
\mathcal{L}_{\text{smooth}} &= \frac{\sum_i (\|\nabla u\|^2 + \|\nabla v\|^2) \cdot m_i}{\sum_i m_i}, \\
\mathcal{L}_{\text{curl}} &= \frac{\sum_i (\partial_x v - \partial_y u)^2 \cdot m_i}{\sum_i m_i}
\end{align}

\noindent where $I_{\text{warp}}$ and $I_{\text{ref}}$ denote warped and reference intensities, $(u, v)$ are distortion field components, and $m_i$ is the spatial mask. The total loss combines these terms as:

\begin{equation}
\mathcal{L} = w_d \mathcal{L}_{\text{data}} + w_s \mathcal{L}_{\text{smooth}} + w_c \mathcal{L}_{\text{curl}},
\end{equation}

\noindent where weights $w_d$, $w_s$, $w_c$ scale data fidelity, smoothness, and curl-free constraints. In the paraxial approximation, the distortion field maps to the gradient of the lens sag and is therefore irrotational ($\nabla \times \mathbf{d} = 0$). Strong curvatures violate this condition through non-linear geometric effects. We keep the regularization weights small (for example, $w_s, w_c \approx 1e-4$) compared with the data weight ($w_d = 10$). The regularization suppresses incoherent tracking noise but does not prevent the data term from driving local reconstruction near steep regions.

\begin{figure}[htbp]
\centering
\includegraphics[width=\linewidth]{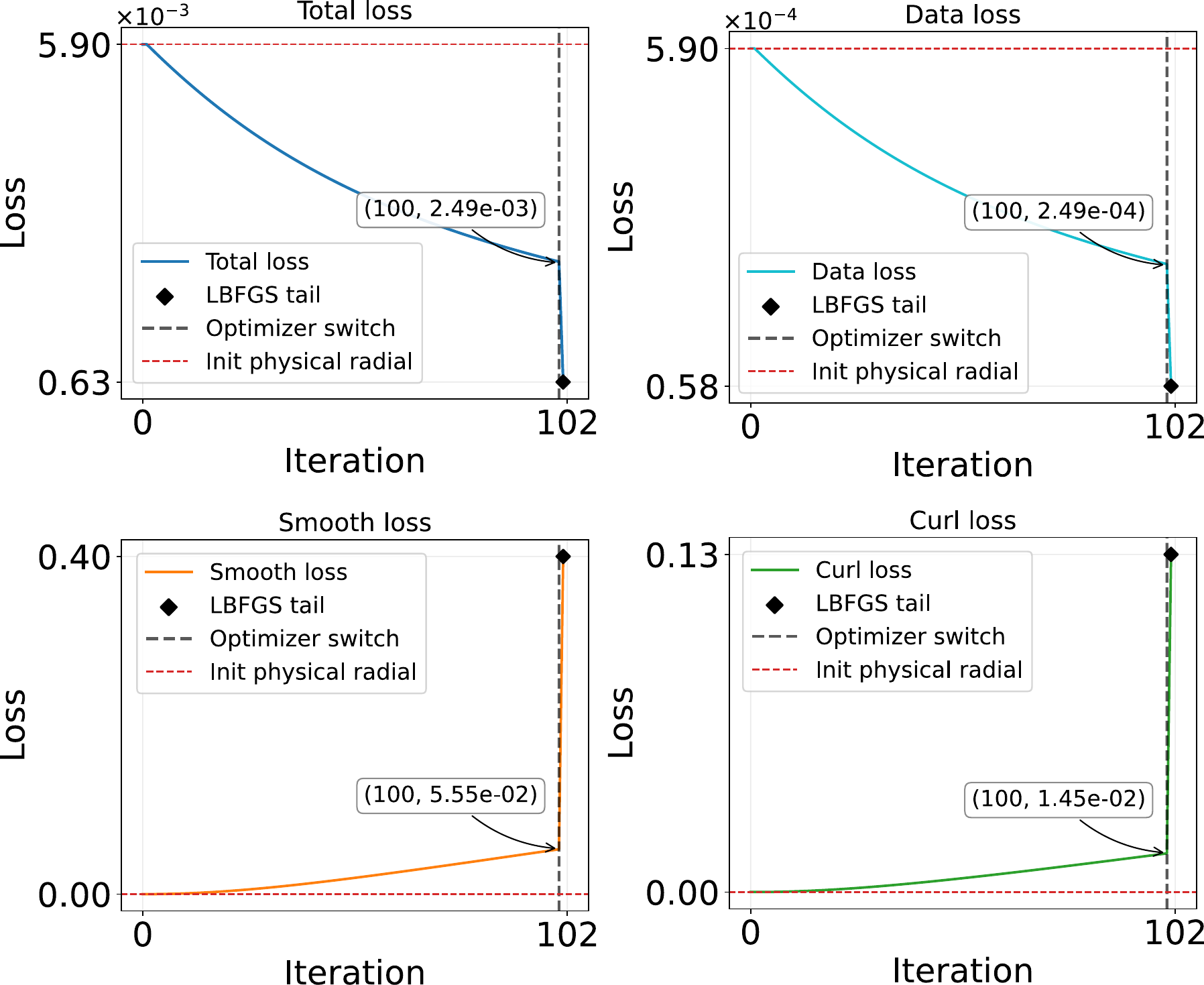}
\caption{Loss history for the LA1608-A distortion-field refinement. The panels show total loss, brightness data loss, smoothness loss, and curl loss. The physical radial initialization provides the starting field, Adam optimization runs for 100 iterations, and the L-BFGS tail starts at the optimizer transition marked at iteration 100.}
\label{fig:supp_fine_tune_loss_history}
\end{figure}

Figure \ref{fig:supp_fine_tune_loss_history} presents the loss history for the LA1608-A distortion-field refinement. The initialization for this run used the physical radial tangent model fitted to sparse feature matches. The subsequent dense refinement used 100 Adam iterations with learning rate \(5\times10^{-3}\), followed by an L-BFGS tail with a 100-iteration maximum and strong-Wolfe line search. For this representative run, the total loss decreased from \(5.90\times10^{-3}\) to \(6.33\times10^{-4}\). The masked brightness MSE decreased from \(5.90\times10^{-4}\) to \(5.81\times10^{-5}\), corresponding to a \(90.2\%\) improvement relative to the initialized field.

The measured runtime for this displacement-retrieval stage was approximately \(3.1\,\mathrm{s}\) on the processing workstation. The radial initialization accounted for \(0.9\,\mathrm{s}\), and dense fine tuning accounted for \(2.1\,\mathrm{s}\). These timings exclude postprocessing ray-tracing-based surface reconstruction.

\section{Forward Ray Model}\label{sec:supp_forward_ray_model}
\subsection{Camera model and non-ideal telecentricity}
\begin{figure}[htbp]
\centering
\includegraphics[width=0.78\linewidth]{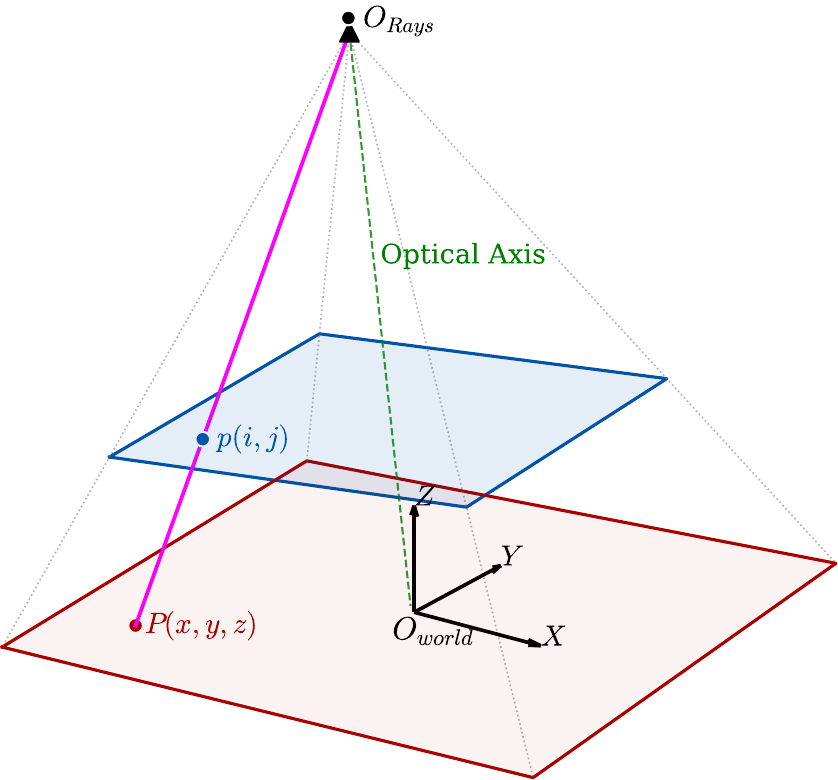}
\caption{Finite-origin projective camera model for near-telecentric imaging; distances have been shortened for clarity. Rays are traced from an effective common origin behind the sensor, cross the virtual sensor plane, and propagate over the working distance \(WD\) toward the world or pattern plane. The calibrated projective ray-origin distance sets the angular spread of the chief rays; in the far-origin limit, the model approaches ideal telecentric projection.}
\label{fig:supp_camera_projection}
\end{figure}

The imaging model is finite-origin (pinhole-like), with optical-axis unit vector
\begin{equation}
\hat{\vv{u}}_{\mathrm{axis}}=(\sin\theta\cos\phi,\sin\theta\sin\phi,\cos\theta),
\end{equation}
and camera center
\begin{equation}
\vv{S}=-d_{\mathrm{origin}}\,\hat{\vv{u}}_{\mathrm{axis}},\qquad d_{\mathrm{origin}}=WD+R_{\mathrm{origin}}.
\end{equation}
Here $WD$ is the working distance (110 mm in the current setup) and $R_{\mathrm{origin}}$ is the calibrated projective ray-origin distance measured from the image plane. Figure~\ref{fig:supp_camera_projection} illustrates this camera geometry.

Coordinates are expressed in millimeters in a camera-fixed world frame. The ray label $\vv{x}$ is the calibrated image coordinate mapped to the aperture-label plane $z=0$, not a physical coordinate on the sensor or mask. For image column $u$ and row $v$, $\vv{g}=\delta(u-u_0,-(v-v_0))$, with $\delta=p_{\mathrm{cam}}/M$ in millimeters per pixel; a measured flow $(f_u,f_v)$ maps to $\vv{d}^{\mathrm{obs}}=\delta(f_u,-f_v)$. The image-row sign is reversed to make world $y$ point upward.

The local incoming direction for grid location $\vv{g}_{ij}^{(3)}=(x_{ij},y_{ij},0)$ is
\begin{equation}
\hat{\vv{v}}_{\mathrm{in}}(\vv{g}_{ij})=\frac{\vv{g}_{ij}^{(3)}-\vv{S}}{\|\vv{g}_{ij}^{(3)}-\vv{S}\|}.
\end{equation}

\paragraph{Calibration vs reconstruction optimization.}
The finite-origin parameter $R_{\mathrm{origin}}$ is derived from telecentricity. For the AL2550 rotational benchmark, the complete camera state, including the axis angles, is fixed; each acquisition has its own lens translation and tilt. The heightmap workflow permits residual pose refinement.

\subsection{Surface intersections and Snell refraction}
Rays are traced through front and back interfaces. For curved interfaces, intersection is solved by Newton iterations \cite{Gross2005}. For a planar interface, a closed-form ray-plane intersection is used.

At each interface, vector Snell refraction is
\begin{equation}
\hat{\vv{v}}_r=\frac{n_1}{n_2}\hat{\vv{v}}_i+\left(\frac{n_1}{n_2}\cos\theta_i-\cos\theta_t\right)\hat{\vv{n}},
\end{equation}
with
\begin{align}
\cos\theta_i&=-\hat{\vv{v}}_i\cdot\hat{\vv{n}},\\
\cos\theta_t&=\sqrt{1-\left(\frac{n_1}{n_2}\right)^2\left(1-\cos^2\theta_i\right)}.
\end{align}
In total internal reflection, the term inside the square root becomes negative and the ray weight is set to zero.

\subsection{Reference and lensed landing positions}
For each aperture-label coordinate $\vv{x}$, the lensed landing point on the comparison plane is $\vv{P}^{\mathrm{final}}_{\theta}(\vv{x})$. The reference (no-lens) landing point is obtained by propagating the unrefracted ray to the same comparison plane $z=z_{\mathrm{cmp}}$,
\begin{equation}
\vv{P}^{\mathrm{ref}}_{\theta}(\vv{x})
=
\vv{S}
+
\frac{z_{\mathrm{cmp}}-S_z}{\hat{v}_{\mathrm{in},z}(\vv{x})}
\hat{\vv{v}}_{\mathrm{in}}(\vv{x}).
\end{equation}
In the baseline AL2550 configuration, $z_{\mathrm{cmp}}$ is the global-$z$ plane through the posed rear vertex. In the slab-assisted configuration, it is the exit plane of the $3.181\,\mathrm{mm}$ slab, whose entrance is anchored at that rear-vertex height. The slab faces remain parallel to the global $x$--$y$ plane. Propagation from the tilted rear surface to the comparison or slab-entrance plane can be signed. The bare reference ray contains neither the lens nor the added slab. These planes follow the optical-stack convention used by the saved reconstruction; they need not coincide with $z=0$.
\section{Predicted Displacement Calculation}
Sensor measurements provide an observed displacement field in image coordinates. Let $\vv{g}_{ij}$ be the reference image-grid coordinate in aperture-label millimeters and let $\vv{x}$ denote a continuous lensed-image coordinate. The ray tracer computes a forward map from an aperture-label coordinate to a physical landing point on the stack-dependent comparison plane,
\begin{equation}
\mathcal{F}_{\theta}(\vv{x})
=
\left[\vv{P}^{\mathrm{final}}_{\theta}(\vv{x})\right]^{xy},
\end{equation}
where $\theta$ denotes the current lens and calibration parameters. The no-lens reference ray gives the corresponding reference landing point $\left[\vv{P}^{\mathrm{ref}}_{\theta}(\vv{g}_{ij})\right]^{xy}$.

This forward map is not itself the optical-flow displacement. Optical flow compares the reference and lensed images in the sensor domain: the pattern feature seen at reference coordinate $\vv{g}_{ij}$ appears at a lensed coordinate $\vv{x}^{*}_{ij}$. Therefore $\vv{x}^{*}_{ij}$ is defined by the inverse correspondence
\begin{equation}
\mathcal{F}_{\theta}(\vv{x}^{*}_{ij})
=
\left[\vv{P}^{\mathrm{ref}}_{\theta}(\vv{g}_{ij})\right]^{xy}.
\label{eq:supp_inverse_condition}
\end{equation}
The predicted sensor-domain displacement is the displacement associated with this inverse coordinate, not the raw difference of two landing points evaluated at the same grid sample.

\subsection{Inverse Correspondence for Sensor-Domain Displacement}\label{sec:supp_inverse_correspondence}
Define the entry displacement field
\begin{equation}
\vv{\Delta}_{\theta}(\vv{x})
=
\left[\vv{P}^{\mathrm{ref}}_{\theta}(\vv{x})\right]^{xy}
-
\left[\vv{P}^{\mathrm{final}}_{\theta}(\vv{x})\right]^{xy}.
\label{eq:supp_entry_delta}
\end{equation}
For on-axis orthographic rays, $\left[\vv{P}^{\mathrm{ref}}_{\theta}(\vv{x})\right]^{xy}=\vv{x}$, and Eq.~\eqref{eq:supp_inverse_condition} becomes the fixed-point relation
\begin{equation}
\vv{x}^{*}_{ij}
=
\vv{g}_{ij}
+
\vv{\Delta}_{\theta}(\vv{x}^{*}_{ij}).
\end{equation}
The iteration is damped in practice,
\begin{equation}
\vv{x}^{(n+1)}
=
(1-\alpha)\vv{x}^{(n)}
+
\alpha\left[
\vv{g}_{ij}
+
\vv{\Delta}_{\theta}(\vv{x}^{(n)})
\right],
\end{equation}
with $\vv{x}^{(0)}=\vv{g}_{ij}$. The sensor-coordinate displacement, expressed in aperture-label millimeters, is the difference between the inverse and reference coordinates:
\begin{equation}
\vv{d}^{\mathrm{pred}}(\vv{g}_{ij})
=
\vv{x}^{*}_{ij}-\vv{g}_{ij}.
\label{eq:supp_predicted_displacement}
\end{equation}

For finite-origin cameras, the reference landing point is not generally equal to the sensor-grid coordinate. Define the projective correction
\begin{equation}
\vv{s}_{\theta}(\vv{x})
=
\left[\vv{P}^{\mathrm{ref}}_{\theta}(\vv{x})\right]^{xy}
-
\vv{x}.
\end{equation}
Then the inverse condition becomes
\begin{equation}
\vv{x}^{*}_{ij}
=
\vv{g}_{ij}
+
\vv{\Delta}_{\theta}(\vv{x}^{*}_{ij})
+
\left[
\vv{s}_{\theta}(\vv{g}_{ij})
-
\vv{s}_{\theta}(\vv{x}^{*}_{ij})
\right].
\label{eq:supp_projective_inverse}
\end{equation}
Thus $\vv{d}^{\mathrm{pred}}=\vv{\Delta}_{\theta}(\vv{x}^{*})+\vv{s}_{\theta}(\vv{g})-\vv{s}_{\theta}(\vv{x}^{*})$. For tilted orthographic rays, the reference offset is constant and cancels. For finite-origin rays it varies across the field. The inverse uses 32 damped iterations with $\alpha=0.5$ and returns the sensor-coordinate displacement $\vv{x}^{*}-\vv{g}$.

\paragraph{Alternative direct correspondence loss.}
The measured optical flow can also be used directly as a correspondence. If $\vv{d}^{\mathrm{obs}}(\vv{g})$ maps reference coordinates to lensed-image coordinates, then the observed lensed coordinate is $\vv{g}+\vv{d}^{\mathrm{obs}}(\vv{g})$. A direct pattern-plane residual would enforce
\begin{equation}
\mathcal{F}_{\theta}\!\left(\vv{g}+\vv{d}^{\mathrm{obs}}(\vv{g})\right)
\approx
\left[\vv{P}^{\mathrm{ref}}_{\theta}(\vv{g})\right]^{xy}.
\label{eq:supp_direct_correspondence}
\end{equation}
This avoids explicitly solving the inverse map because the measured flow supplies the correspondence. For a locally invertible forward map on the same valid support, Eq.~\eqref{eq:supp_direct_correspondence} and the sensor-coordinate residual in Eq.~\eqref{eq:supp_predicted_displacement} have the same zero-residual correspondence. Their least-squares objectives are generally different: locally, the pattern-plane residual is the sensor-coordinate residual multiplied by the forward-map Jacobian. Equal weighting in the two planes therefore does not give an equivalent fit in the presence of nonzero residuals.

\section{Reconstruction Objective and Optimization Stages}\label{sec:supp_reconstruction_objective}
\subsection{Loss}
The reconstruction objective is
\begin{equation}
\mathcal{L}=\mathcal{L}_{\mathrm{data}}+\mathcal{L}_{\mathrm{penalty}},
\end{equation}
where the data term compares predicted and observed displacement on the fixed valid support and
\begin{equation}
\mathcal{L}_{\mathrm{penalty}}=\sum_k\lambda_k\mathcal{P}_k.
\end{equation}
For the AL2550 benchmark, the data term is the component-wise squared L2 objective used by the reconstruction code. For measurement $m$ and component $c\in\{x,y\}$, let $e_{m,c}=d^{\mathrm{pred}}_{m,c}-d^{\mathrm{obs}}_{m,c}$ and let $q_{m,c}^{99}$ denote the 99th-percentile clamp applied to the component-wise squared residual field before support averaging. The reported objective is
\begin{equation}
\mathcal{L}_{\mathrm{data}}=
\frac{1}{\sum_m w_m}
\sum_m w_m
\frac{1}{2}
\sum_{c\in\{x,y\}}
\operatorname{mean}_{\mathcal{S}_m}
\left[
\min\!\left(e_{m,c}^2,q_{m,c}^{99}\right)
\right],
\label{eq:supp_al2550_data_loss}
\end{equation}
with fixed support $\mathcal{S}_m$ for each measurement. The same objective is used for the additional-optimization and displacement-scale tests in Section~\ref{sec:supp_al2550_validation}.

For the general reconstruction pipeline, geometric penalties can be enabled for quantities such as center thickness and edge thickness. In the AL2550 additional-optimization tests discussed below, the lens thickness is fixed and no center-thickness or edge-thickness penalties are active.

\subsection{Three-stage protocol}
\paragraph{Stage 1: conic-only fit.}
The first stage excludes aspheric coefficients and uses
\begin{equation}
z(r)=\frac{r^2}{R\left[1+\sqrt{1-(1+k)\frac{r^2}{R^2}}\right]}.
\end{equation}

\paragraph{Stage 2: heightmap refinement.}
After stage-1 convergence, the single-sided spherical measurements refine the front-surface heightmap and hold the back surface at its fitted parametric shape. Both refracting surfaces contribute to the predicted displacement. Paired front/back measurements can refine two heightmaps. Smoothness regularization remains weak; the data term dominates.

Refinement selects the evaluated state with the lowest data-plus-smoothness objective among states with valid physical ray intersections and stable correspondence inversion. In the fully weighted interior of the finite measurement mask, 32- and 64-step damped inversions must close within $0.01\,\mu\mathrm{m}$ and their queries must agree within $0.01\,\mu\mathrm{m}$. A positive-definite symmetric Jacobian of the bilinear landing map on a convex disk containing all queries supplies a sufficient injectivity check. These numerical checks determine which evaluated states are eligible for restoration. The exported solution is also checked over the full modeled physical ray-label domain. The checks establish numerical consistency, not measurement accuracy.

\paragraph{Stage 3: final parametric fitting.}
The optimized heightmap is reduced to optical shape parameters by fitting a posed parametric surface. The fit uses finite points within the fixed circular aperture in measurement coordinates, with support held constant during pose optimization. This registration step separates surface form from residual translation and orientation, as required for dense 3D measurements without an independent datum \cite{MCBRIDE1996597,HILL2002457,elhayek2014,El-Hayek2015-qe}. The fit estimates vertex-coordinate offsets \((\Delta x,\Delta y,\Delta z)\), measured with respect to the reconstruction coordinate origin, and angular pose \((\alpha,\beta)\). In the implemented routine, the posed surface is obtained from the local sag surface using \(R_{\mathrm{pose}}=R_z(\alpha)R_y(\beta)\). Here, \(\beta\) represents the tilt of the surface axis away from the reconstruction \(z\)-axis. The angle \(\alpha\) specifies the azimuthal direction of that tilt in the \(x\)-\(y\) plane.

After this registration, the local sag follows the ISO 10110-12 asphere convention \cite{ISO10110-12-2019}:
\begin{equation}
z(r)=\frac{r^2}{R\left[1+\sqrt{1-(1+k)\frac{r^2}{R^2}}\right]}+\sum_{i=1}^{N}A_{2i}r^{2i},
\end{equation}
with fitted parameter vector
\begin{equation}
\Theta_{\mathrm{fit}}=\{R,k,A_{2i},\alpha,\beta,\Delta x,\Delta y,\Delta z\},
\end{equation}
obtained by minimizing vertical residuals between the recovered heightmap and the posed model. The optimization uses Adam followed by L-BFGS refinement \cite{kingma2017adam,Liu1989}. For spherical analysis, \(k\) and \(A_{2i}\) are fixed to zero; for aspheric analysis, they are estimated after the heightmap stage, up to \(A_{10}\) in the present implementation. Including \((\alpha,\beta)\) in the fit prevents residual surface tilt from being absorbed into curvature or aspheric coefficients.

The conic--heightmap--parametric sequence above describes the general reconstruction and the heightmap-based residual analyses presented in this Supplement. For the quantitative AL2550 benchmark reported in the main text and in Section~\ref{sec:supp_al2550_validation}, a shared normalized rotational-asphere model with active $R,k,A_4,A_6,A_8$ and fixed $A_{10}=0$ is optimized directly in staged pose, radius, conic, and aspheric refinements on a fixed physical aperture. This keeps the surface comparison tied directly to the rotational-asphere parameters used in the forward model. The AL2550 stages use Adam with a reduce-on-plateau schedule and restore the best state of each stage.

The AL2550 acquisition used the Grasshopper3 camera in Mono16 mode, with 10 effective ADC bits and unsigned 16-bit TIFF storage. Ten accepted frames were averaged with floating-point accumulation for each reference and measurement burst, and each mean was stored as a float32 TIFF. Each physical placement combines two undisturbed repeat bursts; their means are averaged arithmetically before one flow estimate is computed. They share one pose. Seven baseline placements and three slab-assisted placements have equal measurement weight. Each uses its capture-associated reference burst mean; closing reference bursts are saved for drift diagnostics. The fixed $950\times950$ flow crop is sensor-centered. Flow is estimated on a 405-pixel-radius mask, then scored on a fixed $20\,\mathrm{mm}$-diameter disk centered on the paired fiducial-derived carrier center estimate and intersected with valid observed support.

The flow pair is jointly scaled to uint8 and histogram-matched. A physical radial fit initializes dense brightness refinement with data/smoothness/curl weights $10/0.01/0.01$. Adam uses at most 10,000 iterations, initial learning rate 0.005, plateau patience 50, linear learning-rate reduction 0.01 to a minimum $10^{-7}$, and early stopping after 100 iterations without improvement greater than $10^{-7}$. A strong-Wolfe L-BFGS tail has a 100-iteration maximum. No preblur or postfilter smoothing is enabled. Local displacement outliers use a 51-pixel window and 2.5-standard-deviation threshold.

Shared shape starts at $R=25.56\,\mathrm{mm}$, $k=A_4=A_6=A_8=A_{10}=0$, with no random initialization jitter. Initial acquisition poses are restored from the preceding pose-only bootstrap; configured axial rotations remain fixed. Lateral translation and two lens tilt angles remain active at every stage; axial translation, axial rotation, rear-plane shape, thickness, refractive index, and camera state are fixed. Thickness and edge-thickness penalties are disabled.

The five Adam stages are pose only (2500 iterations), radius and pose (1200), conic and pose (1200), and two all-active asphere/pose stages (4000 and 3000). All saved stages completed their budgets. The respective no-improvement stopping patiences are 600, 500, 500, 600, and 600 iterations, with zero minimum improvement. Reduce-on-plateau halves the learning rate after 200, 200, 200, 250, and 300 stagnant iterations, with relative threshold $10^{-7}$ and minimum learning rate $10^{-9}$. Warmup applies to the first and final two stages (300, 100, and 100 steps). Shape and pose coordinates have separate learning rates recorded in the frozen configuration. Each stage restores the state with the lowest optical objective; CCP agreement is evaluated after reconstruction and does not enter the loss or state selection.

The frozen AL2550 camera uses $\theta\approx0.77^\circ$, $\phi\approx90.8^\circ$, telecentricity parameter $0.150^\circ$, $WD=110\,\mathrm{mm}$, and a $22\,\mathrm{mm}$ calibration diameter. The corresponding distances are $d_{\mathrm{origin}}\approx4.21\,\mathrm{m}$ and $R_{\mathrm{origin}}\approx4.10\,\mathrm{m}$. These values are rounded for presentation; the reconstruction configuration stores the full numerical state. Here $\theta$ is measured from $+z$ and $\phi$ from $+x$ toward $+y$; lens pose uses $R_zR_yR_x$. This camera configuration is held fixed during the AL2550 inverse. The calibration table separately summarizes the repeated slab fits.

\section{Residual and Aberration Analysis}\label{sec:supp_residual_analysis}
The optimized heightmap contains spatial information from the data-consistency term. Conventional geometric parameters are obtained only in the subsequent fitting step. After final fitting, residuals are evaluated in the fitted local surface coordinates and inside the clear-aperture mask. Waviness is separated from high-frequency content with an isotropic Gaussian filter \cite{ISO2025}. The Gaussian low-pass convention has 50\% amplitude transmission at the stated cutoff wavelength $\lambda_c$, with $\sigma=\lambda_c\sqrt{\ln2}/(\sqrt{2}\pi)$ in physical coordinates. The example below uses $\lambda_c=1.5\,\mathrm{mm}$ to suppress shorter-period content before Zernike analysis; the cutoff is a processing choice, not a demonstrated optical-performance boundary. The filtered residual is decomposed into Zernike modes to summarize the dominant aberration-like departures \cite{Mahajan2007Zernike}.

\begin{figure}[htbp]
\centering
\includegraphics[width=1.0\linewidth]{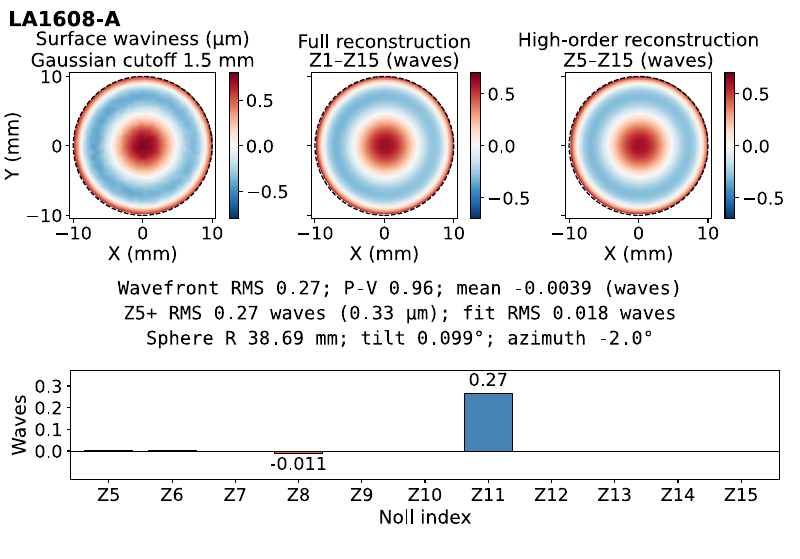}

\caption{Residual and Zernike analysis for the LA1608-A plano-convex running example. The optimized heightmap is randomly sampled at 100,000 valid points (seed 2026), cropped to $20\times20\,\mathrm{mm}$, and fitted with a posed sphere over a fixed 20 mm circular aperture. The conic constant and aspheric coefficients are fixed to zero, leaving non-spherical departures in the residual. Surface waviness is obtained with a 1.5 mm Gaussian cutoff and converted to wavefront error at $\lambda=633\,\mathrm{nm}$ using $n=1.518$. Full reconstructions include $Z_1$--$Z_{15}$; high-order maps and signed coefficient bars show $Z_5$--$Z_{15}$. Wavefront RMS is mean-centered; fit RMS includes the reconstruction-difference mean. All maps share the displayed spatial coordinates, and the two reconstructed maps use the same color limits.}
\label{fig:heightmap_residual_zernike_analysis}
\end{figure}

We consider two residual definitions. The first compares the optimized heightmap \(Z_{\mathrm{hm}}\) with its best-fitting parametric surface \(Z_{\mathrm{fit}}\):
\begin{equation}
\Delta Z_{\mathrm{fit}}(\mathbf{x}) = Z_{\mathrm{hm}}(\mathbf{x}) - Z_{\mathrm{fit}}(\mathbf{x}).
\end{equation}
This residual isolates departures that remain after removing the best-fit global shape and is therefore suited to analyzing local surface structure. The second compares the optimized heightmap with the nominal design surface \(Z_{\mathrm{design}}\):
\begin{equation}
\Delta Z_{\mathrm{design}}(\mathbf{x}) = Z_{\mathrm{hm}}(\mathbf{x}) - Z_{\mathrm{design}}(\mathbf{x}).
\end{equation}
This residual includes both local surface departures and global parameter deviations from design. Either residual can be converted to wavefront error before Zernike decomposition:
\begin{equation}
W(\mathbf{x}) = \frac{n-1}{\lambda}\Delta Z(\mathbf{x}),
\end{equation}
Here $W$ is in waves, $n$ is dimensionless, and $\Delta Z$ and $\lambda$ must use the same length units. This is a thin-phase, near-normal-incidence optical-path approximation for a sag departure replacing air with lens material. It neglects oblique propagation, coupled departures of both surfaces, and downstream wavefront propagation. We use $\lambda=633\,\mathrm{nm}$ as an analysis reference; the AL2550 residual figure uses the fixed index $n=1.518$, so its wave conversion is illustrative and is separate from the 565-nm reconstruction. A wavelength-specific index is required for a dispersive optical prediction. The interpretation depends on the residual definition. \( \Delta Z_{\mathrm{fit}} \) emphasizes residual form error after removing the best-fitting posed surface. \( \Delta Z_{\mathrm{design}} \) measures departure from the design prescription and therefore includes parameter bias, including residual power error. In the present analysis, the residual is fitted with 15 Zernike terms, which is sufficient for the low-order content shown here. The reported high-order reconstruction excludes the first four Noll terms, corresponding to piston, tip/tilt, and defocus, and therefore emphasizes astigmatism, coma, and spherical-aberration-like terms. Higher-order expansions can be used when the residual contains more complex structure \cite{Rimmer:75,Svechnikov:15}. Figure~\ref{fig:heightmap_residual_zernike_analysis} reports \( \Delta Z_{\mathrm{fit}} \) for the LA1608-A example. The design-referenced residual is omitted for compactness.

\begin{figure}[htbp]
\centering
\includegraphics[width=\linewidth]{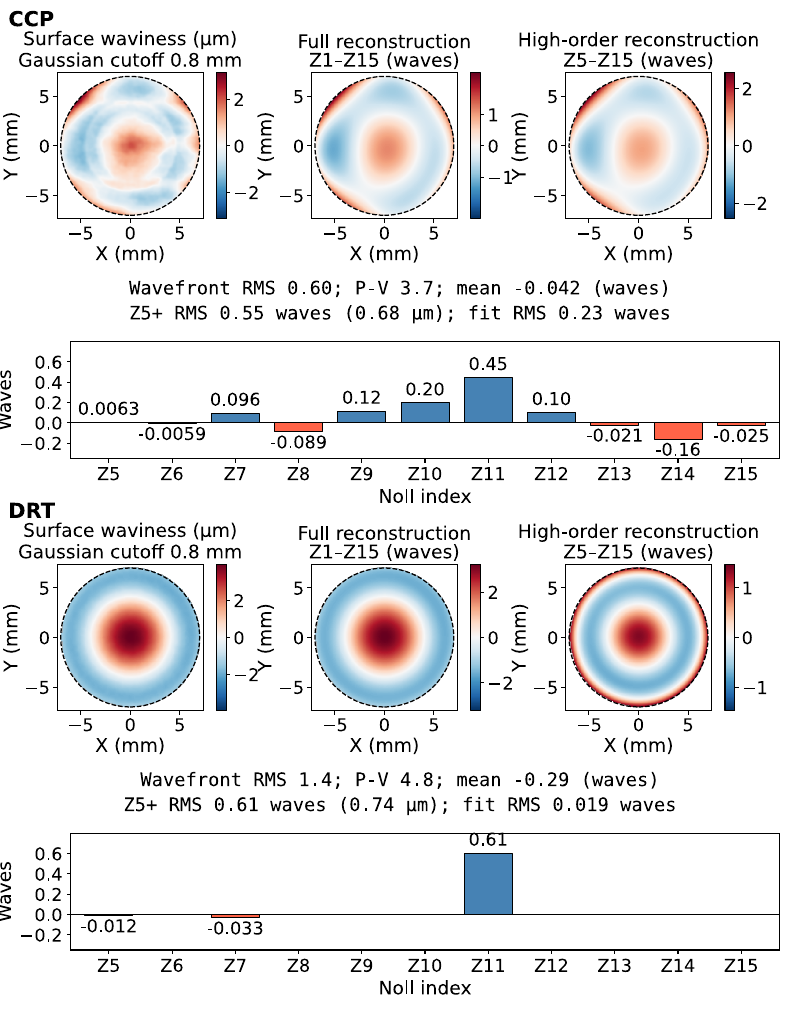}
\caption{Exploratory residual and Zernike comparison for the AL2550 lens over the \(14\,\mathrm{mm}\) circular window set by the usable CCP field. Top: chromatic confocal profilometry (CCP). Bottom: differentiable ray tracing (DRT). A second-order polynomial fitted on a $14\times14\,\mathrm{mm}$ crop is subtracted before filtering with a 0.8 mm Gaussian cutoff. Waviness is converted to wavefront error using \(\lambda=633\,\mathrm{nm}\) and $n=1.518$. Full reconstructions include $Z_1$--$Z_{15}$; high-order maps and coefficient bars show $Z_5$--$Z_{15}$. Wavefront RMS is mean-centered; fit RMS includes the reconstruction-difference mean. Map color scales are independent, and coefficient charts share a common vertical scale. The qualitative modal resemblance is separate from the quantitative rotational-surface benchmark.}
\label{fig:al2550_residual_zernike_comparison}
\end{figure}

We also show an exploratory residual-structure comparison to illustrate the type of spatial information captured by the heightmap reconstruction. This comparison is deliberately kept separate from the quantitative AL2550 benchmark above. The quantitative AL2550 result uses a direct rotational-asphere reconstruction. The residual maps below come from the more flexible heightmap reconstruction, for which pose--form separation and modal amplitudes have not been validated to the same level. Each dataset is randomly subsampled to 100,000 valid height points (seed 2026) and cropped to $14\times14\,\mathrm{mm}$ before fitting and subtracting a second-order 2D polynomial. The nominal aspheric profile is not used for this subtraction. This subtraction removes quadratic structure, including curvature and astigmatism-like terms, regardless of whether it originates in the lens, alignment, or instrument. It does not identify the removed terms as CCP error or the remaining structure as intrinsic lens waviness. The residual is then filtered with a 0.8 mm Gaussian cutoff and converted to wavefront error at \(\lambda=633\,\mathrm{nm}\). The usable CCP field limits the comparison to a circular \(14\,\mathrm{mm}\) window. Figure~\ref{fig:al2550_residual_zernike_comparison} shows that the dominant DRT residual resembles the low-order CCP structure, particularly in the spherical Z11 aberration. Rounded to two significant figures, the Z11 coefficients are 0.45 waves for CCP and 0.61 waves for DRT; the Z5+ coefficient RMS values are 0.55 and 0.61 waves, respectively. These modal summaries are larger for DRT and do not support attenuation of its overall residual amplitude. The maps use independent color limits, so color intensity cannot be compared directly across rows. The coefficient charts share a common vertical scale. At this stage, the apparent modal resemblance is treated only as a qualitative observation. In particular, pose--surface coupling, image-calibration residuals, and representation-dependent smoothing can all alter the recovered residual field. We therefore do not use the Zernike coefficients or the DRT--CCP residual resemblance as a quantitative surface-accuracy validation. Higher camera resolution and improved image-distortion calibration should make this comparison more quantitative.

The aperture limit in Fig.~\ref{fig:al2550_residual_zernike_comparison} comes from the CCP benchmark, not from the transmission geometry itself. The transmission setup is also not constrained by probe standoff in the same way as CCP; in the present instrument, the measurement distance is set by the telecentric working distance, \(WD=110\,\mathrm{mm}\). DRT can use a wider portion of the lens aperture because stronger curvature produces a larger displacement signal and improves conditioning relative to flatter surfaces. CCP and interferometric measurements face the opposite trend: steep slopes reduce the accessible aperture. The advantage has a boundary. Once the refracted rays exceed the angular acceptance of the telecentric imaging system, tracking and reconstruction become unreliable. This limit is related to the telecentricity and ray-acceptance constraints discussed for vision-ray metrology by Ramirez-Andrade and Falaggis \cite{ramirez2024}.

\section{Imaging-System Calibration}\label{sec:supp_camera_calibration}

Imaging-system calibration uses the same differential measurement principle as lens reconstruction. Conventional camera calibration estimates pose and distortion from planar targets observed over several views \cite{Zhang2000}; telecentric variants usually keep that target-based structure and replace pinhole projection with orthographic or near-orthographic models \cite{LI2013538,Duan:22}. Here, the patterned mask remains the target, but the lens sample is replaced by a commercial glass slab of known thickness and measured refractive index. The slab distorts the pattern behind it in a predictable way. Because the displacement is small, the dense Farnebäck optical-flow estimate is sufficient. The differentiable ray-tracing model is then fit to this field to estimate the angular pose of the imaging axis and the finite-origin parameter used for non-ideal telecentricity.

We represent the near-telecentric camera with a finite-origin projective model \cite[Chap.~6]{Hartley2003}. The sensor is located at working distance \(WD\) from the pattern plane, and rays are traced from an effective origin behind the sensor along the calibrated optical axis; in the far-origin limit, this model approaches telecentric projection. Recent telecentric calibration work explicitly models imaging errors and resolves pose-sign ambiguity using constrained joint optimization \cite{ZONG2026109854}; that work does not establish the finite-origin model used here. The calibrated quantities are the angular pose of the telecentric axis, parameterized by \(\theta\) and \(\phi\), and the projective ray-origin distance. From this finite-origin geometry we report the derived maximum chief-ray angle across the aperture as the telecentricity error. Image coordinates are transformed using the calibrated magnification factor \(M = 0.185\times\), taken from the manufacturer technical report.

Experimentally, we capture the usual reference image of the undistorted pattern and a second image with flat commercial glass slabs placed on the pattern. For an ideally telecentric system normal to the pattern plane, a flat slab would produce no lateral distortion after registration. The measured displacement field therefore constrains the angular pose and finite-origin parameter directly. Once estimated, these parameters are reused as long as the relative camera-pattern geometry remains fixed. Conversely, if the camera pose is fixed and the slab thickness is known, the same refractive calibration experiment could be used to estimate the refractive index of an unknown transparent material at the measurement wavelength.

A single-pixel sensitivity estimate follows by assuming that the smallest reliable displacement is one tenth of a camera pixel:
\begin{equation}
    \theta_{\mathrm{min}} \approx
    \frac{0.1p}{\left(1 - n_{\mathrm{air}}/n_{\mathrm{slab}}\right)Mt},
    \label{eq:min_detectable_angle}
\end{equation}
where $p$ is the camera pixel pitch, $M$ is the telecentric magnification, $n_{\mathrm{air}}$ and $n_{\mathrm{slab}}$ are the refractive indices of air and the calibration slab, and $t$ is the slab thickness. Using $p=5.5\,\mu\mathrm{m}$, $M=0.184$, $n_{\mathrm{air}}=1.0003$, $n_{\mathrm{slab}}\approx1.52$, and $t=10\,\mathrm{mm}$ gives $\theta_{\mathrm{min}}\approx0.05^\circ$.

The estimate is conservative because it treats the measurement as if it depended on one local displacement value. In the calibration fit, the parameters are inferred from a dense displacement field over the full aperture; coherent angular changes are constrained by many pixels at once, so the effective repeatability can be better than the single-pixel estimate.

\begin{figure}[htbp]
\centering
\includegraphics[width=\linewidth]{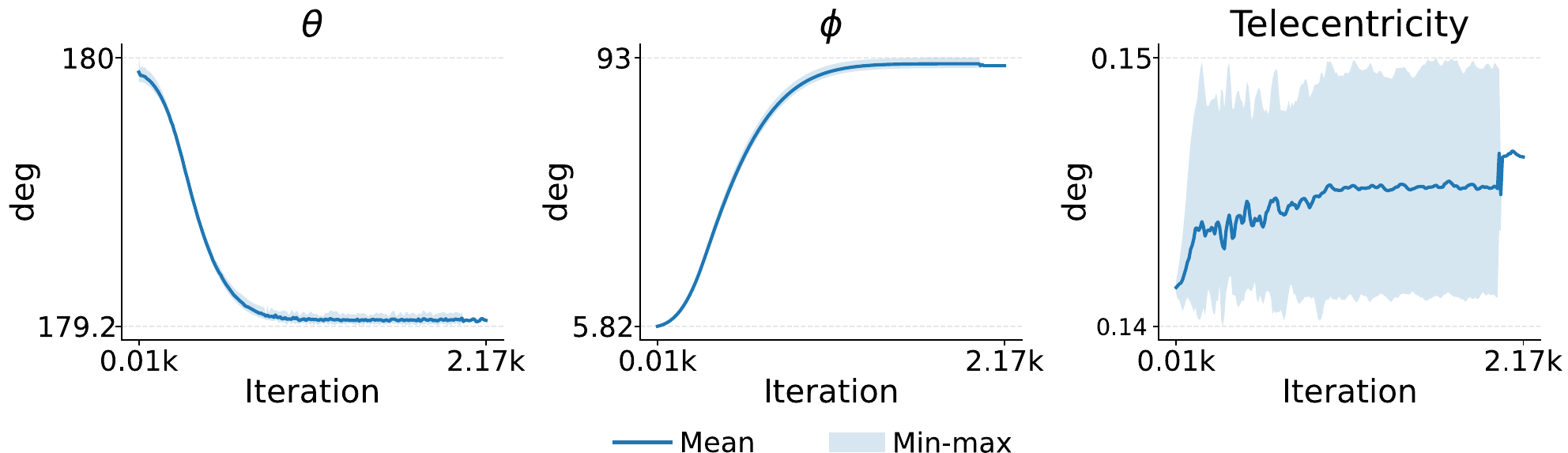}
\caption{Calibration convergence for four repeated slab measurements. Curves show the mean fitted polar angle $\theta$, azimuthal angle $\phi$, and derived telecentricity angle; shaded bands show the min--max range of fits available at each iteration. Fewer than four fits contribute after the first run ends. The four terminal telecentricity values span approximately $0.011^\circ$.}
\label{fig:calibration_convergence}
\end{figure}

The calibration summary uses four independent slab captures. The calibration coordinate frame traces rays toward negative $z$, so the reported polar angle near $179.23^\circ$ corresponds to a tilt of about $0.77^\circ$ from the pattern normal. Reflecting the $z$ axis gives the forward-model convention $\theta_{+z}=180^\circ-\theta_{-z}$; the azimuth is unchanged by this reflection. Each capture was optimized separately; the table summarizes their fitted values. The frozen AL2550 camera state is specified separately in Section~\ref{sec:supp_reconstruction_objective}. Figure~\ref{fig:calibration_convergence} illustrates convergence of the polar angle $\theta$, azimuthal angle $\phi$, and derived telecentricity angle for a separate series of four slab fits. Their terminal telecentricity values span approximately $0.011^\circ$; the plotted envelope uses the fits available at each iteration.

For the measurements reported in the main manuscript, calibration used two 6.5 mm commercial glass slabs stacked on the pattern to increase the signal-to-noise ratio of the retrieved displacement field. Because this material was not compatible with nominal fused quartz or fused silica, its refractive index was measured by spectroscopic ellipsometry and used empirically in the calibration model. The value used in the calibration was \(n=1.4822\) at 565 nm; measurement details are given in Section~\textit{Experimental Refractive-Index Measurements}.

\begin{figure}[htbp]
\centering
\includegraphics[width=1.0\linewidth]{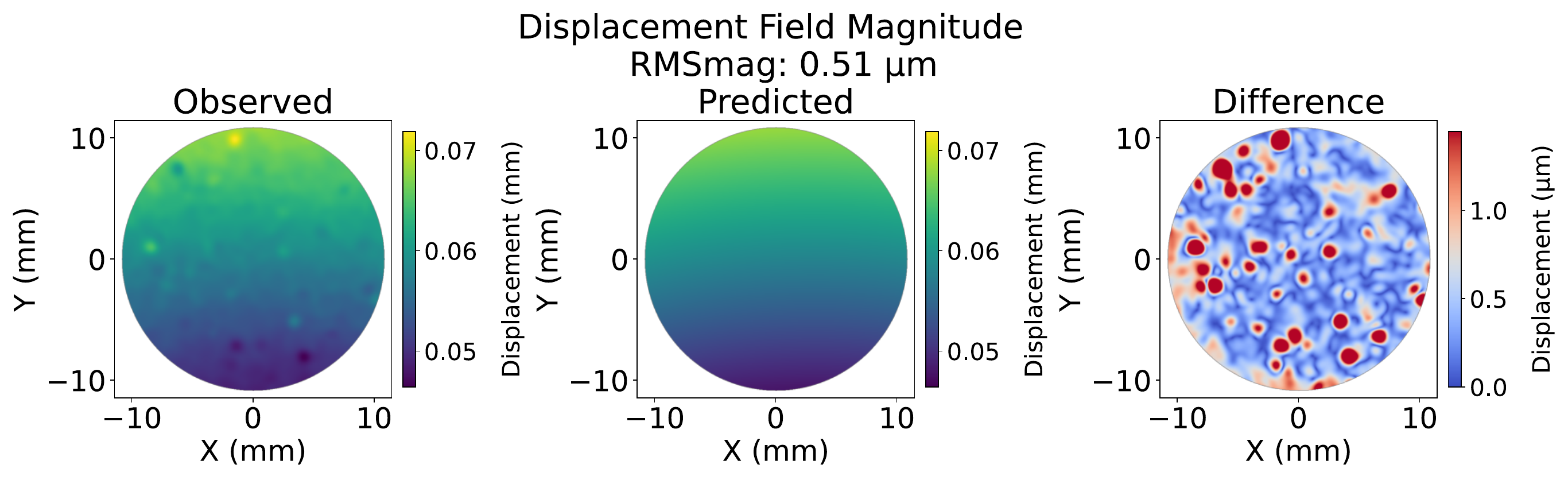}
\caption{Imaging-system calibration using a 13 mm commercial-glass stack composed of two 6.5 mm thick slabs. The measurement estimates the angular pose between the imaging axis and the pattern-plane normal together with the projective ray-origin distance of the near-telecentric camera model. Left: observed displacement-field magnitude. Center: predicted displacement-field magnitude after optimizing angular pose and projective ray-origin distance. Right: absolute displacement residual.}
\label{fig:system_calibration}
\end{figure}

Figure~\ref{fig:system_calibration} shows one calibration measurement, and Table~\ref{tab:angular_measurements} summarizes four independent repeats. The fitted alignment is stable across repeats and shows a small measurable departure from ideal telecentricity. The maximum chief-ray angle fitted by the model is approximately \(0.144^\circ\). This value is the effective telecentricity error of the projective camera model, not a direct remeasurement of the manufacturer's telecentric-lens specification. The effective origin lies approximately \(4.27\,\mathrm{m}\) behind the sensor and \(4.38\,\mathrm{m}\) from the pattern plane, including the \(110\,\mathrm{mm}\) working distance. The residual displacement map in Fig.~\ref{fig:system_calibration} contains localized mid-spatial-frequency structure from the slab measurement and from the smoothing behavior of the Farnebäck optical-flow estimate. A global pose and finite-origin model cannot reproduce all of this structure, but the overall displacement field is fitted with a sub-micrometer RMS residual.

\begin{table}[ht]
\centering
\caption{Calibration summary for four repeated measurements. SD uses divisor 4; the counts denote included/total samples.}
\label{tab:angular_measurements}
\begin{tabular}{lcc}
\toprule
Parameter & Mean $\pm$ SD & $N$ \\
\midrule
$\theta$ ($^\circ$)
    & $179.23 \pm 0.01$ & 4/4 \\
$\phi$ ($^\circ$)
    & $91.1 \pm 1.2$ & 4/4 \\
Telecentricity ($^\circ$)
    & $0.144 \pm 0.004$ & 4/4 \\
Origin-to-sensor distance
    & $4.27 \pm 0.12$ m & 4/4 \\
RMS residual ($\mu$m)
    & $0.60 \pm 0.05$ & 4/4 \\
\bottomrule
\end{tabular}
\end{table}

The present calibration does not yet include full field-dependent image-distortion correction. Its contribution to the reported geometric errors has not been isolated. Unmodeled image distortion can affect both fitted surface parameters and residual-based aberration estimates, so agreement with catalog or CCP surfaces does not establish a bound on this contribution. The dense-field measurement offers a direct path to such a correction: after fitting the known slab refraction and global pose, repeatable residual displacement can be parameterized as imaging distortion. This extension is especially relevant given the close link between telecentricity, wavefront quality, and low-uncertainty optical metrology \cite{Xu:2019}.

\section{Experimental Refractive-Index Measurements}\label{sec:supp_refractive_index_measurements}
\begin{figure}[htbp]
\centering
\includegraphics[width=\linewidth]{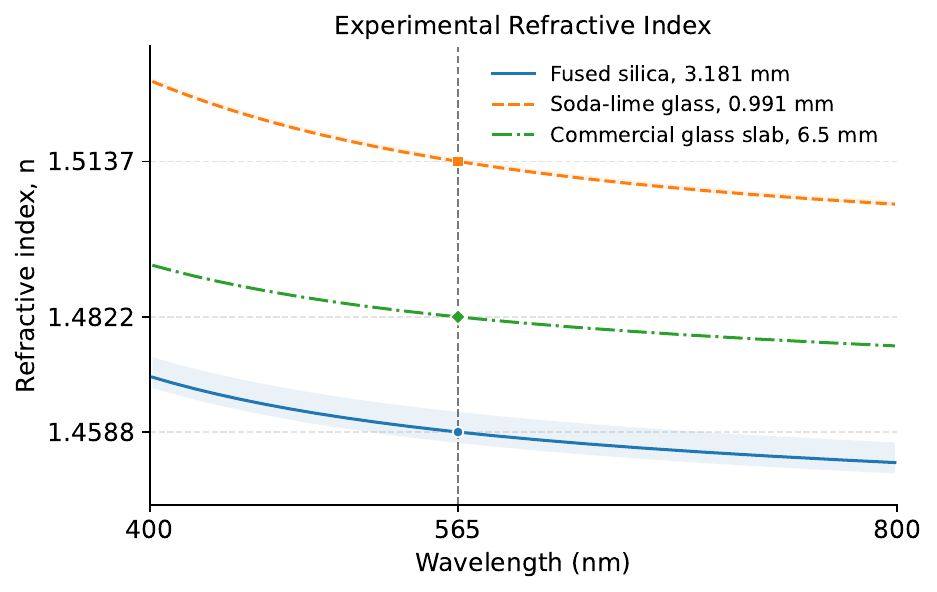}
\caption{Experimental refractive-index dispersion measured with a Woollam M-2000 spectroscopic ellipsometer. The comparison includes fused silica ($3.181\,\mathrm{mm}$, Chemglass Life Sciences CQG-0620-05), soda-lime glass ($0.991\,\mathrm{mm}$ microscope slide), and the commercial glass slab used for calibration ($6.5\,\mathrm{mm}$). The vertical dashed line marks the illumination wavelength used in the calibration model, $565\,\mathrm{nm}$. Interpolation of the mean curves gives $n=1.4588$, $n=1.5137$, and $n=1.4822$, respectively. The commercial glass value is used for camera calibration; the fused-silica and soda-lime values are used when evaluating intermediate layers for two-position measurements.}
\label{fig:refractive_index_measurements}
\end{figure}

The refractive indices of the transparent slabs used in calibration and two-position measurements were measured by spectroscopic ellipsometry; nominal material values were not used. Measurements were performed with a Woollam M-2000 variable-angle spectroscopic ellipsometer, and the values used in the reconstruction were obtained by interpolating the mean measured dispersion curves at the illumination wavelength, $565\,\mathrm{nm}$. Figure~\ref{fig:refractive_index_measurements} shows the measured dispersion curves and their min--max bands.

The camera-calibration measurements used two $6.5\,\mathrm{mm}$ commercial glass slabs stacked on the mask. These slabs were treated as commercial glass after their measured dispersion was found to be incompatible with nominal fused quartz or fused silica. The calibration value was $n=1.4822$ at $565\,\mathrm{nm}$. The same material-characterization procedure was also applied to a $0.991\,\mathrm{mm}$ soda-lime microscope slide and a $3.181\,\mathrm{mm}$ fused-silica slab (Chemglass Life Sciences CQG-0620-05), which were tested as intermediate layers for the two-position measurement scheme.  Interpolation at $565\,\mathrm{nm}$ gave $n=1.5137$ for the soda-lime glass and $n=1.4588$ for the fused silica. The narrow min--max bands indicate low measurement-to-measurement variation in the fitted refractive index.

\section{Measurement Diversity and Conditioning}\label{sec:supp_conditioning_strategies}
Transmission inverse problems can be ill-conditioned because coupled surface and pose changes produce similar transmitted-ray displacements. For AL2550, the rear surface is planar and held known, so the relevant ambiguity is between the correlated rotational-asphere parameters and acquisition pose over the finite aperture. Multiple rotations, reseats, and slab-assisted acquisitions provide complementary sampling and optical sensitivity. In-plane rotation alone does not change the intrinsic shape of a rotational surface; differences between these acquisitions arise from placement, sampling, and optical configuration.

The data contain seven baseline acquisitions (side A) and three slab-assisted acquisitions (side B), with a calibrated \(3.181\,\mathrm{mm}\) fused-silica slab of refractive index 1.4588 inserted between mask and lens. Both sides include nominal \(0^\circ\), \(45^\circ\), and \(90^\circ\) states; side A also includes four nominal-zero reseats.

\begin{figure}[htbp]
\centering
\includegraphics[width=\linewidth]{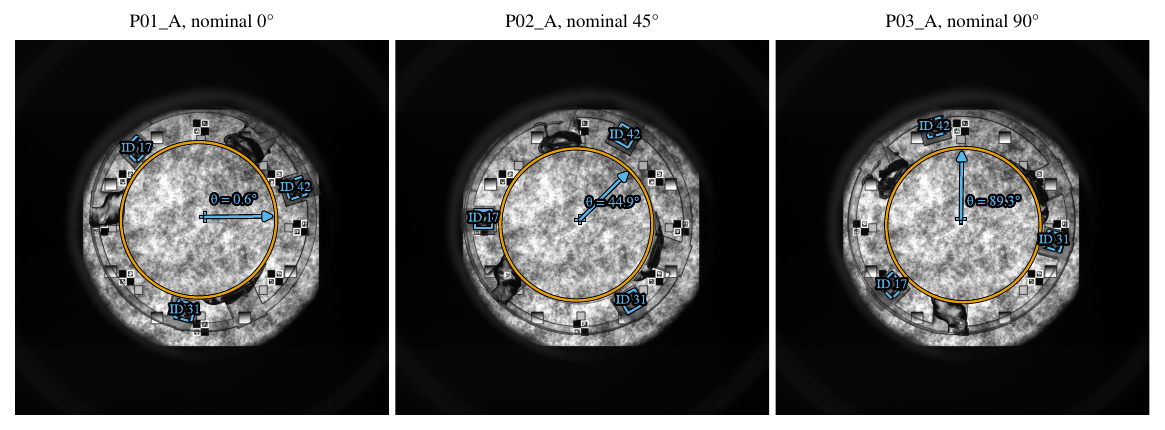}
\caption{Representative AL2550 acquisitions at nominal \(0^\circ\), \(45^\circ\), and \(90^\circ\) in-plane rotation states. Marker labels show tracked coded fiducials, and circles show physical-boundary detections from fixed-reference differential images. These boundary overlays are distinct from the paired fiducial-derived carrier centers used for reconstruction support.}
\label{fig:supp_fiducial_rotation_states}
\end{figure}

A UV-laser-cut transparent tracking ring with coded fiducials provides carrier center estimates and available in-plane orientation measurements. The fixed \(20\,\mathrm{mm}\)-diameter scoring disks use paired fiducial-derived carrier centers, distinct from the detected physical-boundary overlays in Fig.~\ref{fig:supp_fiducial_rotation_states}. Tracking angles are available for eight acquisitions; the rotational reconstruction retains nominal in-plane angles for all ten. Lateral decenter and two-axis tilt are fitted independently and are not forced to the tracking centers. Section~\ref{sec:supp_center_sensitivity} assesses sensitivity to replacing the scoring centers with boundary-derived estimates.

\begin{figure}[htbp]
\centering
\includegraphics[width=0.82\linewidth]{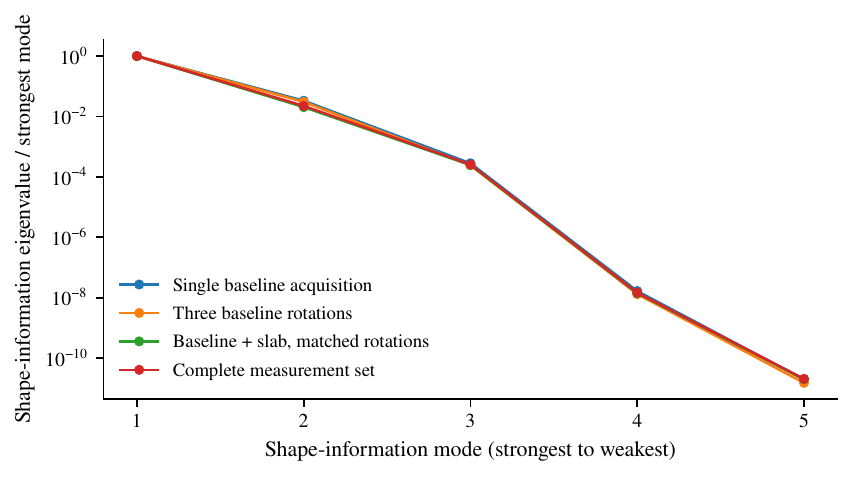}
\caption{AL2550 shape-information spectra after local elimination of per-acquisition pose, all evaluated at the same final ten-acquisition state. The four sets are one baseline acquisition, three baseline rotations, matched baseline and slab-assisted rotations, and all ten acquisitions. Each spectrum is normalized by its largest eigenvalue and shows relative conditioning; weak modes remain separated from the strongest by many orders of magnitude.}
\label{fig:supp_conditioning_spectrum}
\end{figure}

\begin{figure}[htbp]
\centering
\includegraphics[width=0.78\linewidth]{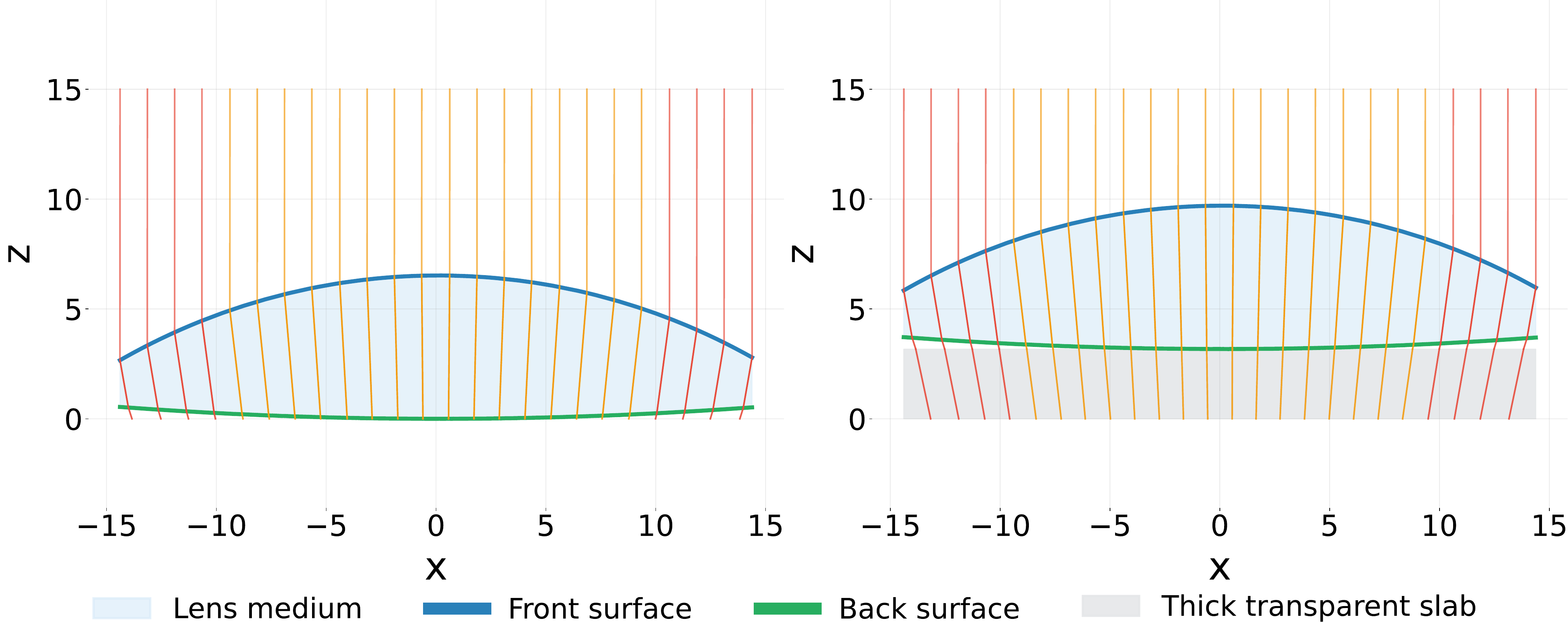}
\caption{Baseline and slab-assisted AL2550 measurement geometries. The baseline acquisition places the lens directly on the patterned mask, whereas the slab-assisted acquisition inserts a calibrated fused-silica layer between mask and lens. These optical configurations are combined with multiple in-plane rotation states and independently optimized acquisition poses to provide measurement diversity.}
\label{fig:supp_conditioning_placeholder}
\end{figure}

All measurement sets are evaluated at the same final ten-acquisition reconstruction, rather than separately optimized shapes. Figure~\ref{fig:supp_conditioning_spectrum} shows the local spectra; Fig.~\ref{fig:supp_conditioning_placeholder} illustrates the two optical geometries.

The Jacobian is computed by central finite differences of the weighted residual representation whose squared norm equals the component-wise L2 objective, including its 99th-percentile squared-error cap. The cap is computed before support selection and recomputed at each perturbation. Locally, \(\delta\mathbf{r}\approx\mathbf{J}_s\delta\mathbf{s}+\mathbf{J}_p\delta\mathbf{p}\), with optimizer shape coordinates \(\mathbf{s}=(R,k,A_{4,n},A_{6,n},A_{8,n})\) and per-acquisition pose coordinates \(\mathbf{p}=(t_x,t_y,r_x,r_y)\). Physical polynomial scales are \(10^{-5}\), \(10^{-9}\), and \(10^{-12}\) for \(A_4,A_6,A_8\). Eliminating the linearized pose increment gives the Schur complement \cite{Triggs2000BundleAdjustment}
\begin{equation}
\mathbf{H}_{\mathrm{shape}}=
\mathbf{J}_{s}^{\mathsf{T}}\mathbf{J}_{s}
-
\mathbf{J}_{s}^{\mathsf{T}}\mathbf{J}_{p}
\left(\mathbf{J}_{p}^{\mathsf{T}}\mathbf{J}_{p}\right)^{+}
\mathbf{J}_{p}^{\mathsf{T}}\mathbf{J}_{s},
\label{eq:supp_profiled_shape_information}
\end{equation}
where \((\cdot)^{+}\) is the Moore--Penrose pseudoinverse \cite{Penrose1955GeneralizedInverse}, using relative cutoff \(10^{-12}\). This accounts for local pose adjustments and is related to projection in separable least squares \cite{GolubPereyra1973}; it is a post-reconstruction diagnostic, not exact optimization over nonlinear poses. Summing the objective-weighted acquisition blocks gives the averaged objective normal matrix. Multiplication by the total acquisition weight gives the reported information sum and preserves condition numbers. We report the normal-matrix positive-spectrum ratio
\begin{equation}
\kappa = \frac{\lambda_{\max}}{\lambda_{\min}^{+}},
\end{equation}
with relative eigenvalue cutoff \(10^{-12}\). The values apply to the stated fixed coordinates; for full rank, this normal-matrix ratio is the square of the effective Jacobian condition number. Each plotted spectrum is divided by its largest eigenvalue, showing relative conditioning rather than absolute sensitivity.

All tested sets retain numerical rank five at this cutoff and remain strongly ill-conditioned. The condition numbers are \(5.1\times10^{10}\) for one baseline acquisition, \(6.6\times10^{10}\) for three baseline rotations, \(6.2\times10^{10}\) after adding nominal-zero reseats, \(5.0\times10^{10}\) for matched baseline and slab-assisted rotations, and \(4.9\times10^{10}\) for all ten acquisitions. The slab modestly improves relative conditioning without resolving coefficient correlation. Added measurements can increase absolute information even when the condition number worsens. These local results are conditional on the fixed calibration, thickness, index, rear surface, plate model, and scoring support; they establish neither global uniqueness nor measurement uncertainty.

\section{CCP Benchmark Processing}\label{sec:supp_ccp_processing}
\begin{figure}[htbp]
\centering
\includegraphics[width=0.78\linewidth]{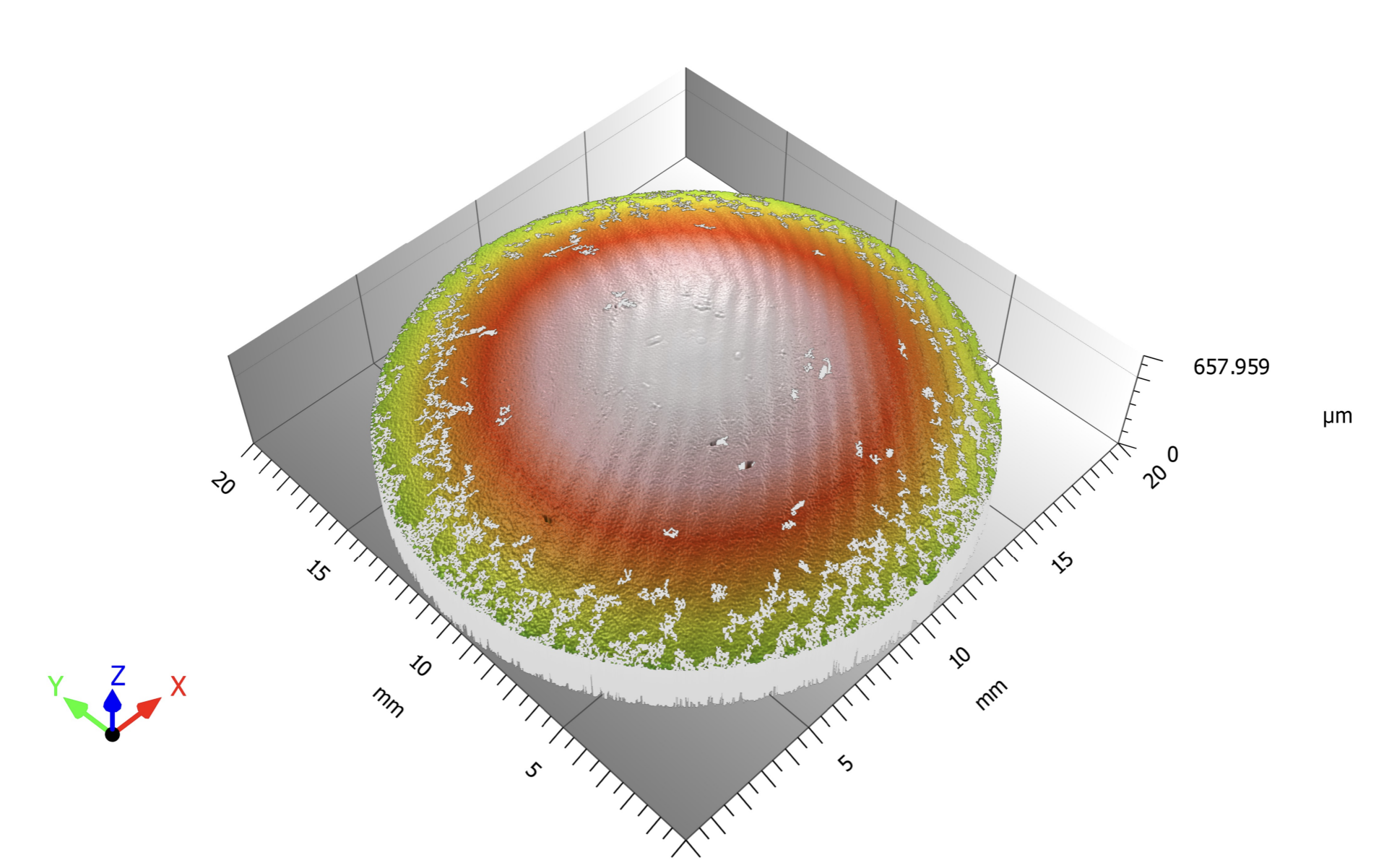}
\caption{3D view of the CCP surface measurement used for the AL2550 benchmark. Quantitative comparison uses the fixed central \(14\,\mathrm{mm}\)-diameter validation window described in the text; the outer exported field is excluded because of the CCP measurement limits.}
\label{fig:supp_ccp_3d_view}
\end{figure}

The lens was measured with a Nanovea ST400 non-contact point-scanning profilometer \cite{nanovea, Blateyron2011, hao2023}. The archived AL2550 XYZ export covers approximately \(22\times22\,\mathrm{mm}\) with \(10\,\mu\mathrm{m}\) lateral sampling and was produced using the thresholded export from the original profilometer processing. Figure~\ref{fig:supp_ccp_3d_view} shows a three-dimensional view of the measured surface. The lateral calibration factors applied to the exported coordinates are 1.0127 and 1.0135 in the two scan directions. The total AL2550 sag is approximately \(3.1\,\mathrm{mm}\), larger than the \(1.1\,\mathrm{mm}\) probe range, so the outer steep region is not used as the quantitative validation aperture. The measurable slope is also limited by the objective numerical aperture \cite{Blateyron2011,nanovea}.

For the benchmark reported in the main text, the CCP validation support is frozen independently of the fitted surface. The aperture center is estimated from the measured apex, after which a fixed \(14\,\mathrm{mm}\)-diameter circular window is selected in the CCP measurement frame. From the archived XYZ export onward, finite samples inside this window are included; no additional fit-dependent point rejection, interpolation, replacement, or iterative support refinement is applied. The same frozen point set is then reused for all catalog, free-asphere, and DRT comparisons. This avoids the support changes that can otherwise occur when the aperture is repeatedly re-centered on a fitted surface.

The manufacturer prescription is first registered to the CCP data with a five-degree-of-freedom rigid alignment. A separate rotational-asphere fit is then performed on the same fixed support with that alignment held fixed. Its free shape parameters are \(R\), \(k\), \(A_4\), \(A_6\), \(A_8\), and \(A_{10}\). The fit is initialized without the manufacturer aspheric coefficients. The CCP measurement is therefore an independent data source, while the rigid alignment used for the rotational fit is inherited from the catalog-registration step. Individual polynomial coefficients are not used as the primary benchmark because the finite-aperture parameterization is strongly correlated.

For two rotational sag profiles \(z_1(r)\) and \(z_2(r)\) over an aperture of radius \(a\), we report the circular-area RMS difference
\begin{equation}
\sigma_{z,A}=
\left[
\frac{2}{a^2}
\int_0^a
\left(z_1(r)-z_2(r)\right)^2 r\,dr
\right]^{1/2}.
\label{eq:supp_area_weighted_rms}
\end{equation}
The factor \(r\) weights each radial interval by annular area, so this quantity is the RMS sag difference over the full circular aperture under rotational symmetry, not an unweighted RMS over uniformly sampled radial coordinates.

The raw CCP point residual to the rigidly aligned manufacturer surface is \(1.21\,\mu\mathrm{m}\) RMS over the fixed support. This quantity includes non-axisymmetric/local structure and measurement residuals. For comparison with the rotational DRT model, we therefore use the circular-area RMS sag difference defined in Eq.~\eqref{eq:supp_area_weighted_rms}. The fitted CCP rotational surface differs from the manufacturer prescription by \(0.09\,\mu\mathrm{m}\) RMS over the central \(14\,\mathrm{mm}\) diameter. The selected DRT surface differs from the manufacturer surface by \(0.50\,\mu\mathrm{m}\) RMS and from the CCP rotational surface by \(0.44\,\mu\mathrm{m}\) RMS over the same comparison region.

\section{AL2550 Reconstruction Validation and Systematic Controls}\label{sec:supp_al2550_validation}
The DRT reconstruction is optimized on fixed \(20\,\mathrm{mm}\)-diameter scoring disks centered on paired fiducial-derived carrier estimates and intersected with valid observed support. These disks remain fixed while lateral decenter and tilt are fitted independently. The joint reconstruction contains seven baseline measurements and three slab-assisted measurements, with a separate acquisition pose for each measurement and one shared rotational lens surface. Side B includes the measured fused-quartz slab in the forward model. The central \(14\,\mathrm{mm}\) region is used only when comparing the resulting rotational surface with CCP; CCP is never used in the reconstruction objective.

\subsection{Stagewise surface trajectory}
\begin{figure}[htbp]
\centering
\includegraphics[width=0.80\linewidth]{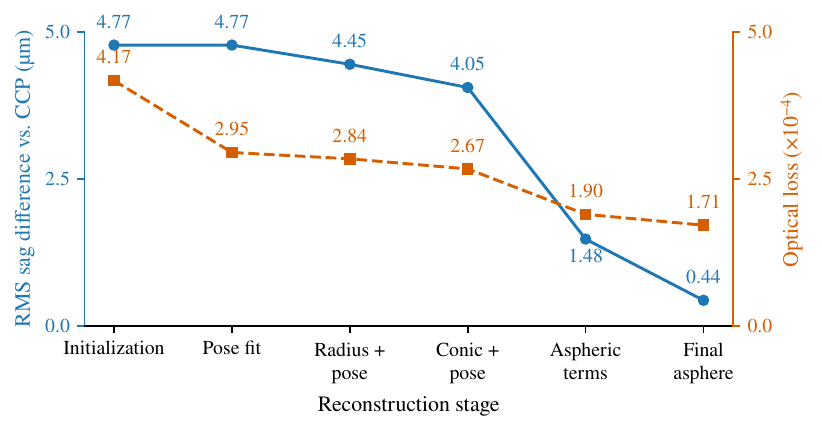}
\caption{AL2550 reconstruction trajectory from initialization through the completed optimization stages. Blue circles, left axis: circular-area RMS sag difference relative to the independent CCP rotational fit over the common central \(14\,\mathrm{mm}\)-diameter aperture, evaluated using Eq.~\eqref{eq:supp_area_weighted_rms}. Orange squares, right axis: optical loss from fitting the measured displacement fields. Both quantities share the stage axis. The initialization and pose-adjusted states have identical surface sag because the first stage optimizes acquisition poses only. CCP is not part of the reconstruction objective.}
\label{fig:supp_al2550_stagewise_ccp}
\end{figure}

Figure~\ref{fig:supp_al2550_stagewise_ccp} shows the optical loss and the circular-area surface difference to the independently acquired CCP rotational surface at initialization and after each reconstruction stage. Pose adjustment reduces the optical loss without changing the initialized surface; subsequent shape optimization reduces both quantities through the final reconstruction reported in the main text.

\subsection{Scoring-center sensitivity}\label{sec:supp_center_sensitivity}
We compare the reported fiducial-derived scoring centers with centers obtained by averaging physical-boundary detections from the two replicate images. Two matched 2000-iteration joint continuations start from the same final shape and poses, using identical frozen displacement fields and optimization settings; only the fixed scoring centers differ. The recovered surfaces differ by \(0.0088\,\mu\mathrm{m}\) area RMS and at most \(0.0213\,\mu\mathrm{m}\) over the central \(14\,\mathrm{mm}\) diameter. Their RMS differences from the CCP rotational reference are \(0.521\,\mu\mathrm{m}\) for fiducial centers and \(0.518\,\mu\mathrm{m}\) for boundary centers. Both exceed the reported reconstruction's \(0.435\,\mu\mathrm{m}\) despite lower optical loss, separating the small center effect from the larger continuation drift. The reported reconstruction is retained.

\subsection{Additional optimization and parameter identifiability}
To test whether the reported reconstruction is limited by incomplete convergence, 2000 additional iterations are run from the exact final state with three parameter sets: pose only, lens shape only, and joint pose plus lens shape. Pose-only optimization reduces the objective by 0.11\% and leaves the sag unchanged. Shape-only optimization reduces the objective by 1.3\% but changes the recovered sag by \(0.40\,\mu\mathrm{m}\) RMS and increases the CCP difference from \(0.44\) to \(0.52\,\mu\mathrm{m}\). Joint pose--shape optimization gives the same surface behavior, with a 1.5\% objective reduction and a \(0.40\,\mu\mathrm{m}\) RMS sag change. Thus, additional minimization of the displacement objective does not necessarily improve the recovered surface.

Independent optimization using only the baseline measurements or only the slab-assisted measurements changes the sag in almost the same radial direction. The corresponding RMS sag changes are \(0.35\) and \(0.41\,\mu\mathrm{m}\), and the two radial changes are nearly collinear. The slab-assisted measurements give a larger reduction of their displacement objective, but the common sag direction shows that the bias is not specific to the slab-assisted geometry.

For the complete ten-measurement reconstruction, the normalized shape normal matrix is strongly ill-conditioned, with a positive-spectrum condition number of approximately \(5\times10^{10}\). Its weakest modes are dominated by the normalized \(A_6\) and \(A_8\) coordinates. Repeating the additional optimization with \(A_6\) and \(A_8\) fixed and only \(R\), \(k\), and \(A_4\) active produces essentially the same sag change while achieving about 98\% of the objective reduction. The same low-order sag direction is obtained from the baseline-only and slab-assisted-only measurements. The observed sag bias therefore does not require the weakly constrained high-order coefficients.

\subsection{Effective displacement-scale diagnostic}
We next profile a multiplicative scale applied only to the frozen measured displacement field \(\mathbf{d}^{\mathrm{obs}}\), keeping the spatial grid, reconstruction aperture, lens state, poses, camera model, and slab model fixed. The profile uses the exact reconstruction objective (component-wise L2 with the reported 99th-percentile clamp). The best shared scale corresponds to a +0.16\% displacement change and reduces the ten-measurement objective by 0.29\% without changing the surface. Separate diagnostic profiles give +0.067\% for side A and +0.24\% for side B.

\begin{figure}[htbp]
\centering
\includegraphics[width=0.92\linewidth]{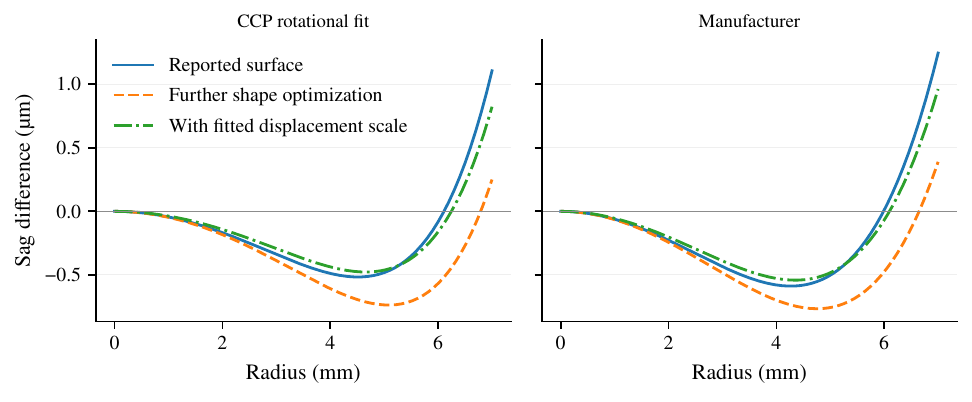}
\caption{Radial sag difference relative to the CCP rotational fit (left) and manufacturer prescription (right) for the selected reconstruction, an extended low-order shape optimization, and the same extended optimization after profiling a shared multiplicative displacement-scale factor. Further shape optimization alone reduces the optical objective but drives the recovered surface away from both independent references. Fitting the displacement-scale factor suppresses most of this additional sag change. The scale factor is used only as a diagnostic and is not interpreted as a calibrated correction to the optical-flow measurement.}
\label{fig:supp_al2550_scale_profile}
\end{figure}

The shared factor is then applied to the frozen displacement fields and the \(R,k,A_4\)-only optimization is repeated from the same reported reconstruction. The scale-only objective and the initial objective of this run agree to numerical precision, confirming the intended initial state. With the displacement-scale adjustment, the subsequent low-order sag change decreases from \(0.40\) to \(0.10\,\mu\mathrm{m}\) RMS, a 75\% reduction. The final surface also agrees more closely with the independent references than the selected reconstruction: the RMS difference to CCP decreases from \(0.44\) to \(0.37\,\mu\mathrm{m}\), and the difference to the manufacturer prescription decreases from \(0.50\) to \(0.43\,\mu\mathrm{m}\). Figure~\ref{fig:supp_al2550_scale_profile} compares the radial sag differences relative to both independent references.

This experiment identifies an effective displacement-scale mode that is causally coupled to the recovered low-order surface. It does not by itself identify the physical source of that mode. A multiplicative displacement change can represent optical-flow scale, spatial calibration, camera geometry, or another forward-model mismatch. The scale factor is therefore used only to diagnose sensitivity to displacement calibration and is not applied as a calibrated correction to the measurement system.

\section{Synthetic Surface Recoverability Controls}\label{sec:supp_synthetic_recoverability}
The AL2550 validation above concerns the rotational asphere recovered from measured displacement fields. Separate synthetic controls were used to determine which spatial-frequency and amplitude ranges are recoverable by the geometric inverse and by the complete image-processing chain. These controls quantify synthetic recoverability and are not experimental resolution measurements. The synthetic aperture has a $25\,\mathrm{mm}$ diameter; $f$ cycles per aperture denotes a directional sinusoid with spatial frequency $f/25\,\mathrm{mm}^{-1}$, independent of orientation. Thus the tested frequencies are 0.08, 0.16, 0.32, and $0.48\,\mathrm{mm}^{-1}$. Surface metrics use a separate $301\times301$ evaluation grid over 95\% of the $12.5\,\mathrm{mm}$ radius; the representative ray-only surface map shows the full aperture. The sampled injected truth uses a $1025\times1025$ heightmap. These apertures differ from the $20\,\mathrm{mm}$ real-data reconstruction and $14\,\mathrm{mm}$ rotational comparison.

\begin{figure}[htbp]
\centering
\includegraphics[width=0.80\linewidth]{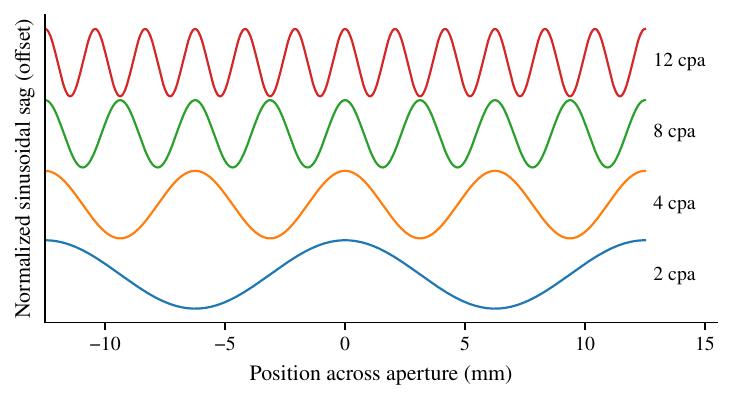}
\caption{Representative directional sinusoidal surface perturbations used in the synthetic recoverability controls. The curves show normalized sag profiles across the aperture for 2, 4, 8, and 12 cycles per aperture.}
\label{fig:supp_recoverability_modes}
\end{figure}

The injected perturbations are directional sinusoidal sag fields spanning 2, 4, 8, and 12 cycles per aperture, with both \(0^\circ\) and \(45^\circ\) orientations and sine/cosine phase. Their RMS sag is defined over the circular aperture after removal of the mean component. Figure~\ref{fig:supp_recoverability_modes} illustrates the spatial-frequency family used in the controlled studies.

\subsection{Ray-tracing surface recovery}
The first synthetic test isolates the ray-tracing inverse before image formation and optical-flow estimation. Six RMS sag amplitudes, \(0.1\), \(0.25\), \(0.5\), \(1\), \(2\), and \(5\,\mu\mathrm{m}\), were evaluated at each spatial frequency. For each frequency--amplitude pair, the four orientation/phase combinations were propagated through the AL2550 ray model for three fixed acquisition poses. Recovery used a 16-mode Cartesian-frequency basis containing all tested sinusoidal modes, and the acquisition pose \((t_x,t_y,r_x,r_y)\) was optimized for each synthetic measurement. This test therefore includes modal cross-coupling and pose uncertainty within the ray-tracing inverse.

For truth and recovered fields $t$ and $r$ on common support, recovery gain is $\langle t,r\rangle/\langle t,t\rangle$, correlation is the centered Pearson correlation, and NRMSE is $\mathrm{RMS}(r-t)/\mathrm{RMS}(t)$. For the ray-only modal inverse, gain is the recovered injected-mode coefficient divided by its injected amplitude; crosstalk is the largest absolute off-diagonal coefficient ratio. Passing requires gain in $[0.90,1.10]$, crosstalk $\leq0.10$, surface NRMSE $\leq0.10$, surface correlation $\geq0.99$, and pose-explainable displacement energy $\leq0.80$. Displacement statistics use a five-pixel stride.

\begin{figure}[htbp]
\centering
\includegraphics[width=0.96\linewidth]{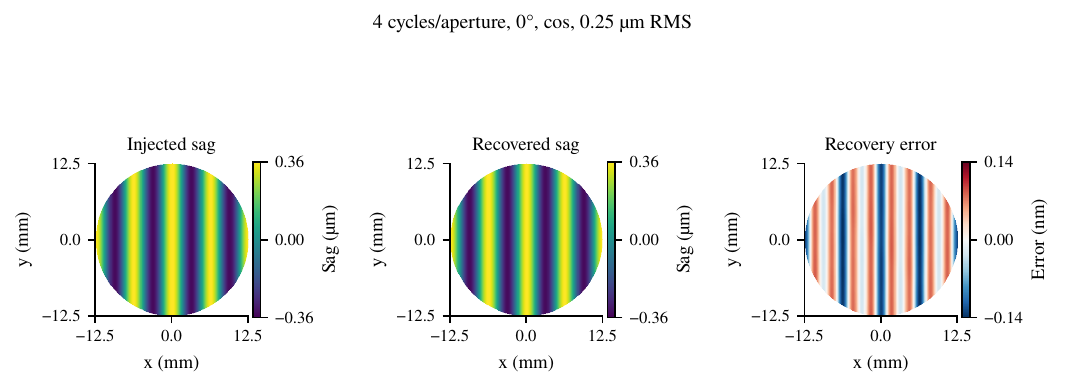}
\caption{Ray-tracing recovery of a 4-cycles-per-aperture, \(0.25\,\mu\mathrm{m}\)-RMS directional sinusoidal sag perturbation. Left: injected sag. Center: recovered sag using the 16-mode sinusoidal basis with acquisition-pose increments fitted simultaneously in the linearized inverse. Right: recovered minus injected sag, shown in nanometers. The recovery-error RMS over the full displayed aperture is \(0.065\,\mathrm{nm}\) (normalized RMS error \(2.6\times10^{-4}\)). Image formation and optical-flow estimation are not included in this test.}
\label{fig:supp_d1c_representative}
\end{figure}

All 24 frequency--amplitude combinations passed the predefined geometric recovery criteria over the tested range. Figure~\ref{fig:supp_d1c_representative} shows a representative 4-cycles-per-aperture, \(0.25\,\mu\mathrm{m}\)-RMS case. This amplitude lies below the recovery limit obtained after adding image formation and optical-flow estimation, showing that the ray-tracing inverse itself is not the limiting step at this spatial frequency.

\subsection{Complete synthetic image-chain recovery}
The second synthetic test evaluates the complete image-processing chain
\[
\text{surface}\rightarrow\text{ray tracing}\rightarrow\text{synthetic images}
\rightarrow\text{optical flow}\rightarrow\text{surface amplitude}.
\]
The same 2, 4, 8, and 12 cycles-per-aperture family was tested over the six sag amplitudes for two matched camera models and two image-formation tiers. The ideal renderer includes resampling and area integration but omits blur, illumination variation, added noise, and capture quantization. The Farnebäck adapter histogram-matches and normalizes the pair, then converts both tiers to 8-bit images internally. The realistic renderer additionally uses a Gaussian PSF with $\sigma=0.7$ sensor pixels, radial illumination rolloff 0.10, horizontal gradient 0.03, independent Gaussian read noise with $\sigma=0.5$ digital counts on an 8-bit scale, and 8-bit capture quantization. Both tiers use twofold supersampling, inverse resampling, and area integration at the original pixel pitch. The source is a captured reference texture reconstructed by Lanczos interpolation and scaled to 0.90 of full intensity; these are controlled synthetic inputs, not independently calibrated image-formation parameters.

The cameras are an orthographic model with parallel tilted chief rays and a matched finite-origin spherical-ray model; both use the same aperture labels, base surface, and pose. The complete-image test uses $950\times950$ images at $0.0297\,\mathrm{mm/pixel}$, with a 329-pixel-radius support. It estimates the known sinusoid amplitude from relative flow between base-lens and perturbed-lens images, after separately removing affine flow components. The Farnebäck settings are pyramid scale 0.5, five levels, 15-pixel window, ten iterations, polynomial neighborhood five, polynomial $\sigma=1.2$, and Gaussian weighting, with no iterative dense refinement.

Each complete-image mode must have at least 90\% common support and a relative-inversion 99th-percentile error $\leq0.01$ pixels. Profiled flow must have gain in $[0.90,1.10]$, correlation $\geq0.95$, and NRMSE $\leq0.30$. Recovered sag must have amplitude gain in $[0.90,1.10]$, NRMSE $\leq0.10$, and correlation $\geq0.99$ after independent piston/tilt removal on the evaluation aperture. Crosstalk is tested by the ray-only 16-mode inverse; the complete-image test fits one known mode. A frequency--amplitude cell passes only if all four orientation/phase combinations pass every criterion; the reported limit is the smallest passing tested amplitude for which every larger tested amplitude also passes. The discrete grid does not establish a continuous detection threshold. Across two cameras and two tiers, the test contains 384 mode records and 96 cells.

The base seed is 2026. Capture seeds are $2027+1000\,\mathrm{round}(1000f)+10j+2c+q$, where $j$ indexes the amplitude/orientation/phase combination from zero, $c$ indexes the camera, and $q$ indexes the image tier. The ideal renderer adds no stochastic noise. Ray-only recovery uses no random sampling.

\begin{figure}[htbp]
\centering
\includegraphics[width=0.76\linewidth]{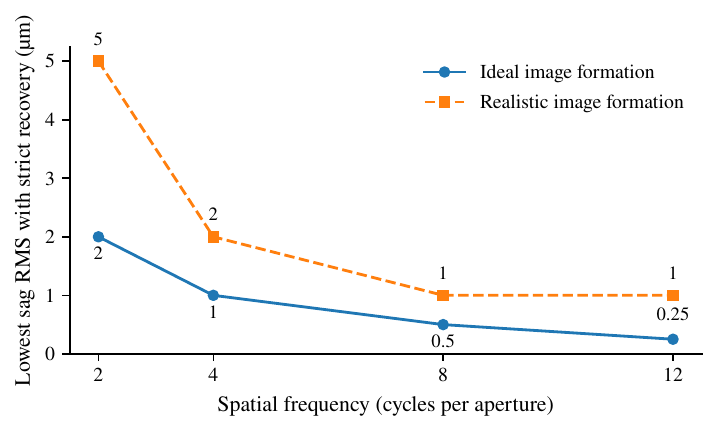}
\caption{Synthetic recovery limit for the complete image and optical-flow pipeline. The realistic image model requires larger sag amplitudes than the ideal image model, particularly at 2--8 cycles per aperture. These limits apply to recovery of the known sinusoidal basis and are not experimental surface-resolution limits.}
\label{fig:supp_f1_recovery_floor}
\end{figure}

The lowest amplitudes satisfying the recovery criteria were \(2\), \(1\), \(0.5\), and \(0.25\,\mu\mathrm{m}\) RMS for ideal image formation and \(5\), \(2\), \(1\), and \(1\,\mu\mathrm{m}\) RMS for realistic image formation at 2, 4, 8, and 12 cycles per aperture, respectively. The two camera models gave the same limits. Figure~\ref{fig:supp_f1_recovery_floor} summarizes these results.

In this test the sinusoidal basis is known and only the perturbation amplitudes are recovered. The resulting frequency--amplitude limits therefore apply to this controlled basis and do not establish general freeform-surface recoverability.

\subsection{Cross-representation controls}
\begin{figure}[htbp]
\centering
\includegraphics[width=0.98\linewidth]{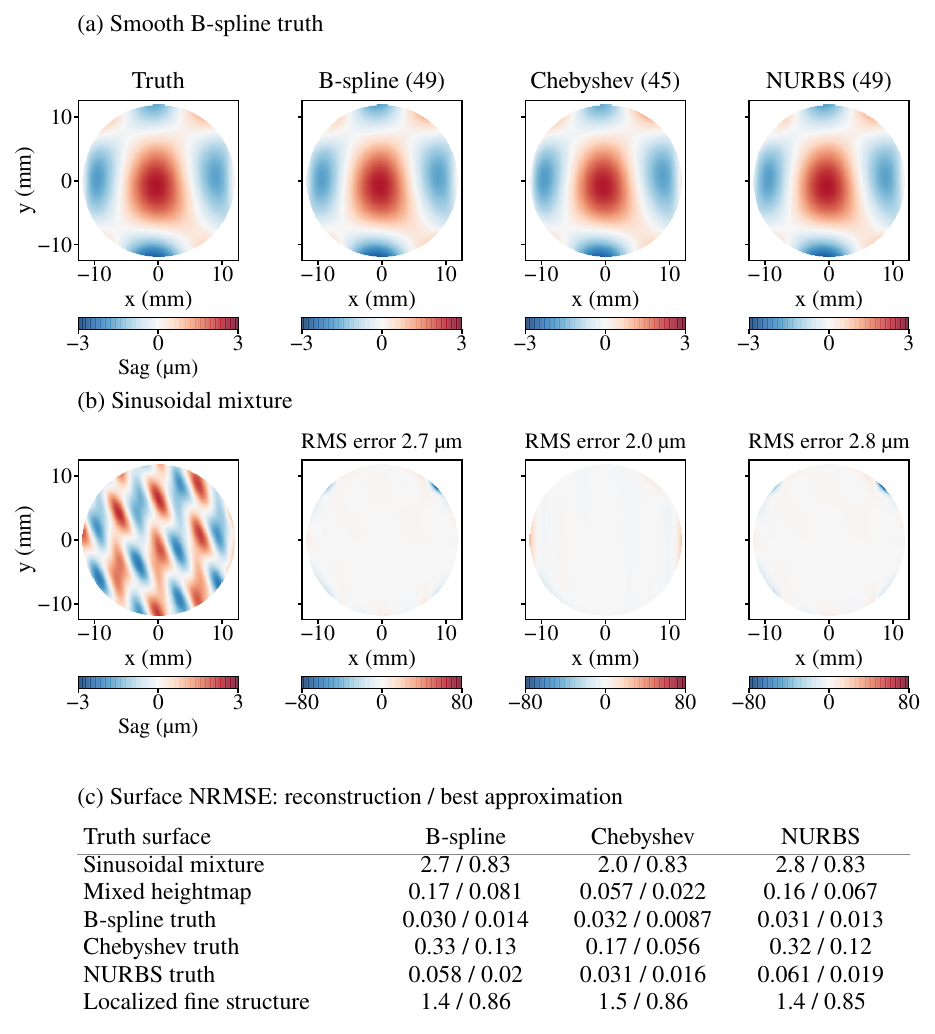}
\caption{Cross-representation ray-only reconstruction. (a) Smooth B-spline truth and its three reconstructions, on a common $\pm3\,\mu\mathrm{m}$ sag scale. (b) Sinusoidal-mixture truth and its reconstructions; the truth uses $\pm3\,\mu\mathrm{m}$ and all three reconstructions use $\pm80\,\mu\mathrm{m}$ to display the full sag excursions without clipping. All maps show the complete evaluation disk. (c) Surface NRMSE for all six truths: each entry gives reconstruction error followed by the best approximation error from a direct least-squares sag fit in the same basis. Both errors allow only piston/tilt alignment and use the same evaluation disk. The approximation fit uses full numerical rank and no ray data; it is a bound on basis flexibility, not a displacement-based reconstruction.}
\label{fig:supp_cross_representation}
\end{figure}

Figure~\ref{fig:supp_cross_representation} compares ray-only recovery of six independent truth surfaces using a cubic $7\times7$ B-spline basis (49 surface parameters), a total-degree-8 Chebyshev basis (45 parameters), and a cubic $7\times7$ NURBS basis (49 parameters, with fixed nonuniform weights). The truths are a directional sinusoidal mixture at 4 and 8 cycles per aperture, a mixed sampled heightmap, higher-resolution B-spline and NURBS surfaces, a degree-10 Chebyshev surface, and a sampled surface with localized fine structure. The truth and recovery models differ in basis or resolution. Each truth has $1\,\mu\mathrm{m}$ RMS sag after piston/tilt removal over the same 95\%-radius evaluation disk used above.

The inverse uses three acquisition geometries with fixed finite-origin camera parameters and fits independent in-plane translations and tilts as alignment parameters. Alternating grid samples are used for fitting and held-out ray evaluation. Surface coordinates are normalized to equal RMS sag, the pose subspace is projected out, and a common relative singular-value cutoff of 0.01 is applied to the linearized surface inverse. Each recovered state is then evaluated with the nonlinear ray model. A comparison passes only if surface NRMSE is $\leq0.10$, surface correlation is $\geq0.99$, held-out ray-perturbation NRMSE is $\leq0.10$, and at least 99.9\% of held-out support remains valid. Image formation and optical-flow estimation are excluded.

All three bases recover the smooth B-spline and NURBS truths under these criteria; only Chebyshev also recovers the mixed heightmap. None recovers every tested truth. The sinusoidal mixture and localized fine structure have best-approximation NRMSE of approximately 0.83--0.86 even when their sag is supplied directly to the surface fit, showing that the tested bases have insufficient flexibility for these cases. Reconstruction errors exceed these approximation bounds. For the degree-10 Chebyshev truth, the Chebyshev inverse gives held-out ray NRMSE of 0.063 but surface NRMSE of 0.17. Agreement with the displacement field therefore does not guarantee accurate recovery of an arbitrary freeform surface. These scores include extrapolation within the evaluation disk beyond parts of the sampled ray footprint.

\clearpage
\bibliography{supplement/references}